\documentclass[11pt]{iopart}
\usepackage{graphics}
\usepackage{iopams}  
\usepackage{amsthm}

\begin{document}

\title{Feedback Control of Traveling Wave Solutions of the
Complex Ginzburg Landau Equation}

\author{K.A.~Montgomery\dag\ and M.~Silber\dag}

\address{\dag\ Department of Engineering Sciences
        and Applied Mathematics, Northwestern University, Evanston, IL
        60208 USA}

\begin{abstract}
  Through a linear stability analysis, we investigate the
  effectiveness of a noninvasive feedback control scheme aimed at
  stabilizing traveling wave solutions $Re^{iKx+i\omega t}$ of the
  one-dimensional complex Ginzburg Landau equation (CGLE) in the
  Benjamin-Feir unstable regime.  The feedback control is a
  generalization of the time-delay method of Pyragas~\cite{Pyr92},
  which was proposed by Lu, Yu and Harrison~\cite{LYH96} in the
  setting of nonlinear optics. It involves both spatial shifts, by the
  wavelength of the targeted traveling wave, and a time delay that
  coincides with the temporal period of the traveling wave.  We derive
  a single necessary and sufficient stability criterion which
  determines whether a traveling wave is stable to all perturbation
  wavenumbers.  This criterion has the benefit that it determines an
  optimal value for the time-delay feedback parameter.  For various
  coefficients in the CGLE we use this algebraic stability criterion
  to numerically determine stable regions in the $(K,\rho)$--parameter
  plane, where $\rho$ is the feedback parameter associated with the
  spatial translation.  We find that the combination of the two
  feedbacks greatly enlarges the parameter regime where stabilization
  is possible, and that the stability regions take the form of
  stability tongues in the $(K,\rho)$--plane.  We discuss possible
  resonance mechanisms that could account for the spacing with $K$ of
  the stability tongues.
\end{abstract}



\pagestyle{myheadings}
\thispagestyle{plain}
\markboth{K. A. MONTGOMERY AND M. SILBER}{FEEDBACK CONTROL OF TRAVELING WAVES}


\submitto{Nonlinearity}

\maketitle

\section{Introduction}
\label{sec:introduction}

\newtheorem{lemma}{Lemma}
\newtheorem{corollary}{Corollary}
\newtheorem{theorem}{Theorem}

A current challenge to our understanding of pattern-forming systems
lies in our ability to control the spatio-temporal chaos that many of
these systems naturally exhibit. The mathematical existence of a
plethora of simple spatial or spatio-temporal patterns for many
nonequilibrium, spatially extended systems is well-established on the
basis of equivariant bifurcation theory~\cite{GSS88}. However, these
simple patterns often prove to be unstable in a given system, which
evolves instead to a state of spatio-temporal chaos.  Perhaps the
simplest and best studied manifestation of spatio-temporal chaos is
that associated with the one-dimensional complex Ginzburg Landau
equation (CGLE)~\cite{SPSHCH92}, a universal amplitude equation that
describes spatially-extended systems in the vicinity of a Hopf
bifurcation.  In the so-called Benjamin-Feir unstable regime, the
simple solutions ({\it e.g.} traveling plane waves and
spatially-homogeneous oscillations) are all unstable to long-wave
perturbations. The focus of this paper is a linear stability analysis
of Benjamin-Feir unstable traveling wave solutions of the CGLE in the
presence of a feedback control scheme that was originally proposed by
Lu, Yu and Harrison~\cite{LYH96} in the setting of nonlinear optics.
The feedback approach is noninvasive, meaning that the feedback signal
decays to zero once the targeted traveling wave state of the system is
realized.

Feedback control methods aimed at stabilizing the unstable periodic
orbits associated with low-dimensional chaotic attractors have been
extensively investigated for more than ten years now, and have proved
especially effective for nonlinear optical systems. The simple
approach taken by Ott, Grebogi and Yorke~\cite{OGY90} relies on
applying small perturbations to a system parameter that help maintain
the system in a neighborhood of the desired unstable periodic orbit.
However, this approach, which requires an active monitoring of the
state of the system so that the feedback is appropriately adjusted,
can prove impractical in systems that evolve too rapidly.
Autoadjusting feedback control methods, in which the feedback is based
upon current and past states of a system, have proven useful for
rapidly evolving systems because they adjust automatically to rapid
changes of the system and require no active monitoring. One
autoadjusting feedback technique that has attracted considerable
attention was introduced by Pyragas \cite{Pyr92}. In this approach the
feedback is proportional to the difference between the current and
past states of a system, {\it i.e.} the feedback is $F=\gamma
(x(t)-x(t-\Delta t))$, where $\Delta t$ is the period of the targeted
unstable periodic orbit.  The method possesses a couple of properties
which make it attractive experimentally. First, if a state with the
desired periodicity is stabilized the feedback term vanishes and
control is achieved in a reasonably noninvasive manner.  Second, this
type of feedback control may be straightforward to implement in the
laboratory when feedback loops are practical.  The method has been
implemented successfully in a variety of experimental systems
including electronic \cite{PT93,GSCS94}, laser \cite{BDG94}, plasma
\cite{PBA96,MKPAPB97,FSK02}, and chemical systems
\cite{SBFHLM93,LFS95,PMRNKG}. A number of modifications of the method
of Pyragas have also been investigated. For instance, Labate~{\it et
al.}~\cite{LCM98} added a filter to their time delayed feedback scheme
for a $CO_2$ laser and successfully stabilized periodic behavior; the
fi lter rejected one of the characteristic frequencies associated with
a quasiperiodic route to chaos in this system. In yet another
direction Socolar~{\it et al.}~\cite{SSG94} proposed a method of
``extended time delay autosynchronization'' which incorporates
information about the state of the system at many earlier times
$t-n\Delta t$ (prescribed by positive integers $n$). This extended
time delay feedback can successfully stabilize traveling wave
solutions of the one-dimensional CGLE~\cite{BS962}, although it fails
in two-dimensions~\cite{HS01}.
   
For spatially extended pattern forming systems, modifications of the
time-delay autosynchronization scheme of Pyragas have been proposed
which take into account not only the temporal periodicity, but also
the spatial periodicity of the targeted pattern.  Perhaps the simplest
such scheme is the one proposed in~\cite{LYH96}, which is the one we
investigate here. Specifically, Lu, Yu and Harrison used numerical
simulations of the two-dimensional Maxwell-Bloch equations describing
a three level laser system to demonstrate that spatio-temporal chaos
(``optical turbulence'') could be eliminated by applying a linear
combination of time-delay feedback and an analogous feedback term
involving a spatial translation, {\it i.e.}, $F=\rho(E({\bf x}+{\bf
  x_0})-E({\bf x}))$, where ${\bf x_0}$ is the translation vector
associated with the feedback. Other schemes for controlling
spatio-temporal patterns have incorporated spatial filters in the
time-delayed feedback.  For instance, Bleich~{\it et al.}~\cite{BHMS97}
showed, through linear stability calculations and numerical
simulations, that traveling wave solutions of a model of a spatially
extended semiconductor laser could be stabilized using a combination
of a Fourier filter and an extended time-delay scheme involving
multiple time delays. Baba~{\it et al.}~\cite{BASJ02} considered
problems for which a Fourier filter is not appropriate; they
incorporate a spatio-temporal filter in their time-delay feedback that
is derived from a linear stability analysis of the targeted unstable
periodic orbit in the uncontrolled problem. Other successful
approaches to feedback control of spatial patterns in nonlinear optics
have relied on the Fourier filters
alone~\cite{DV96,MSOF98,MS98,VS98,BKNT00}.  Numerous investigations of
oscillatory chemical patterns have incorporated a global time-delay
feedback control~\cite{FBS99,KBPvMRE01,BM01,BBMRE03}.  In this case
the magnitude of the global feedback depends on the spatial average of
a chemical concentration at an earlier time.  Global delayed feedback
has been shown to suppress spatio-temporal chaos associated with the
CGLE~\cite{BM96, BPM97}.

The goal of our paper is to provide a detailed analysis of the spatial
and temporal feedback control scheme proposed by Lu, Yu, and Harrison
\cite{LYH96}. In particular our goal is to derive conditions under
which this method can stabilize traveling waves in the Benjamin-Feir
unstable regime of the CGLE. The analysis is complicated because of
the time-delay, and because we must determine stability with respect
to perturbations at all spatial wavenumbers. On the other hand, it is
simplified because the exact traveling wave solutions of the CGLE
(both with and without feedback) take the particularly simple plane
wave form $Re^{iKx+i\omega t}$ (where $R$ and $\omega$ are simple
functions of the wavenumber $K$). Moreover, as is the case for the
CGLE without feedback, the plane wave solution enters the linear
stability analysis only parametrically through the wavenumber $K$.
Specifically, the linear stability analysis reduces to an
investigation of a pair of complex linear delay-differential equations
whose coefficients depend on the wavenumber $K$ of the targeted plane
wave solution and the wavenumber $Q$ of the perturbation but not
explicitly on the spatial variable $x$ (nor on time). Our main linear
stability result is an algebraic condition that determines whether or
not there exists a time-delay feedback term that will stabilize the
traveling wave (for a given spatially-shifted feedback term).
Interestingly, our analysis identifies an {\it optimal} time delay
feedback to use in determining whether the spatial and temporal
feedback control method will stabilize the traveling wave.  Our
algebraic condition for stability, which must be checked numerically,
reveals that the combination of temporal-delays and spatial-shifts in
the feedback greatly enlarges the regions where stabilization is
possible. In some instances we are able to understand the failure of
the feedback control scheme in terms of simple resonant conditions
between the wavenumber $K$ of the underlying traveling wave pattern
and the wavenumber $Q$ of the unstable perturbation, or between the
frequency $\omega$ of the traveling wave and the frequency associated
with the perturbation. These resonances can be tuned to some extent
using the spatially-shifted feedback term and thus are not fundamental
obstacles of the type that can be present when time-delay feedback
alone is used~\cite{JBORB97,N97,NU98}.  (Another approach to
eliminating fundamental limitations of time-delay feedback in
low-dimensional chaotic systems has been proposed and investigated by
Pyragas~\cite{P01}.) We expect our paper to contribute to the body of
literature in which the success of time-delay feedback methods is
investigated through a detailed linear stability analysis of the
associated delay-differential
equations~\cite{JRKFHB00,JBR03,P02,JBORB97,BS96}.

Our paper is organized as follows.  In section~\ref{sec:background} we
describe the feedback to be investigated, and review the basic
properties of traveling wave solutions of the CGLE without feedback.
This section also reviews the stability analysis of a related problem,
in which a time delay feedback term was applied to a system near a
Hopf bifurcation~\cite{RSJ2000}.  Section~\ref{sec:analysis} sets up
our linear stability analysis for the traveling wave solutions of the
CGLE in the presence of feedback. We then prove our main
stability result which allows us to determine the conditions under
which the feedback parameters can be chosen so that Benjamin--Feir
unstable traveling waves are stabilized.  In section
\ref{sec:results}, we apply our stability criterion to numerically
determine the parameter regions in which feedback control is possible,
and discuss possible cause for failure.  Finally, in section
\ref{sec:conclude}, we summarize our results and suggest possible
extensions of our analysis.

\section{Background and Problem Formulation}
\label{sec:background}

The 1--D complex Ginzburg Landau equation, an amplitude equation
describing a spatially extended system near the onset of a Hopf
bifurcation, is given by
\begin{equation}
\partial_{t} A =  A + ( 1 + i b_{1}) \partial _{x}^{2}A 
-(b_{3}-i)|A|^{2}A.  \label{eq:CGLE} \end{equation}
where $A(x,t)$ is a complex amplitude, and $b_1$ and $b_3$ are real
parameters.
We consider traveling wave solutions of~(\ref{eq:CGLE}) of the form
\begin{equation}
A_{TW}=R e^{iKx+i \omega t},
\label{eq:ATW}
\end{equation}  
where the amplitude $R$ and frequency $\omega$ are determined by the 
wavenumber  $K$ through
the relations
\begin{equation}
R^2={(1-K^2)\over b_3}, \label{eq:dispersion1} \end{equation}
\begin{equation}
\omega=R^{2}-b_{1} K^{2}. \label{eq:dispersion2} \end{equation}
We focus on the case of a supercritical Hopf bifurcation so that
$b_3>0$; thus $|K|<1$ follows from~(\ref{eq:dispersion1}). 

\begin{figure}
\centerline{\resizebox{3in}{!}{\includegraphics{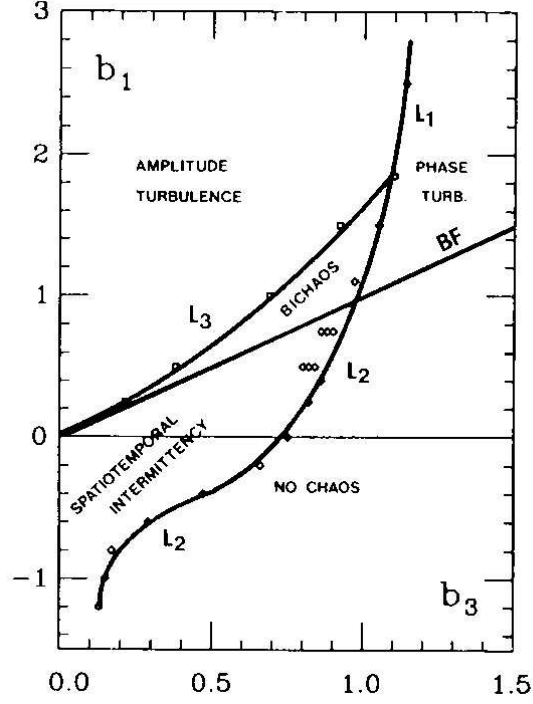}}}
\caption{Phase diagram  for the 1-D complex Ginzburg Landau 
  equation~(\ref{eq:CGLE}) in the $(b_3,b_1)$--parameter plane.  All plane wave
  solutions~(\ref{eq:ATW}) are unstable above the Benjamin--Feir (BF)
  line \(b_{1} = b_{3} \). This figure is reproduced
 (with permission of the author) from~\cite{Chate94}.}
\label{fig:phase}
\end{figure}

In the Benjamin--Feir unstable regime ($b_1>b_3>0$), all plane wave
solutions of the form~(\ref{eq:ATW}) are unstable.  In the
Benjamin--Feir stable regime ($b_{1} < b_{3}$, $b_3>0$), traveling
waves with wavenumbers $K$ satisfying $0 \leq K^{2} < K_{max}^{2}=
(b_{3}-b_{1})/(3b_{3}-b_{1}+2/b_{3})$ are linearly stable to longwave
perturbations~\cite{BF67,SD78,JPBCRK92}.  Figure~\ref{fig:phase}
summarizes the behavior of typical solutions of~(\ref{eq:CGLE}) in the
($b_{1}$, $b_{3}$)--parameter plane. The Benjamin--Feir unstable
regime is divided into three different regions, an amplitude turbulent
regime, a phase turbulent regime, and a bichaotic regime.  The
amplitude turbulent regime is characterized by large fluctuations in
amplitude and phase defects.  In the phase turbulent regime amplitude
fluctuations are much smaller and no defects are observed. In the
bichaotic regime either amplitude or phase turbulence occurs depending
upon initial conditions~\cite{SPSHCH92}.  

We consider how feedback of the type proposed in~\cite{LYH96} 
affects the linear stability of the plane wave solutions 
of the CGLE in the Benjamin--Feir unstable regime. 
In particular, we consider
\begin{equation}\partial_{t}
A =  A + ( 1 + i b_{1}) \partial _{x}^{2}A -(b_{3}-i)|A|^{2}A + F 
\label{eq:mainamp}\end{equation}
where the feedback term, $F$, is a linear combination of spatially
translated and time--delay feedback terms.  Specifically,
\begin{equation} 
F = \rho [A(x+\Delta x,t) - A(x,t)]+\gamma [A(x,t) - A(x,t-\Delta t)],  
\label{eq:spatialfb} \end{equation}
where $\gamma$ and $\rho$ are real parameters and $\Delta t > 0$.
Note that if a solution $A(x,t)$ is spatially periodic with period $
L=\Delta x /n,\ n \in Z $ and temporally periodic with period $
T=\Delta t /m,\ m \in Z $
, the feedback term vanishes.  Thus such a 
space and time periodic state is also a solution of the CGLE 
without feedback.  We will restrict our linear stability 
analysis to traveling plane wave solutions~(\ref{eq:ATW}) 
which have the same spatial and temporal scales as the
feedback,
{\it i.e.} we assume
\begin{equation}
\Delta x = 2 \pi/K\ , \quad \Delta t = 2 \pi/|\omega| \ .
\label{eq:2pi}
\end{equation}  
Note that this choice of spatial feedback scale assumes that the targeted
traveling wave state is not too close to bandcenter ($K=0$). The
temporal feedback assumes $\omega\ne 0$, or equivalently
$K^2\ne 1/(1+b_1b_3)$.

\begin{figure}
\centerline{\resizebox{3.5 in}{!}{\includegraphics{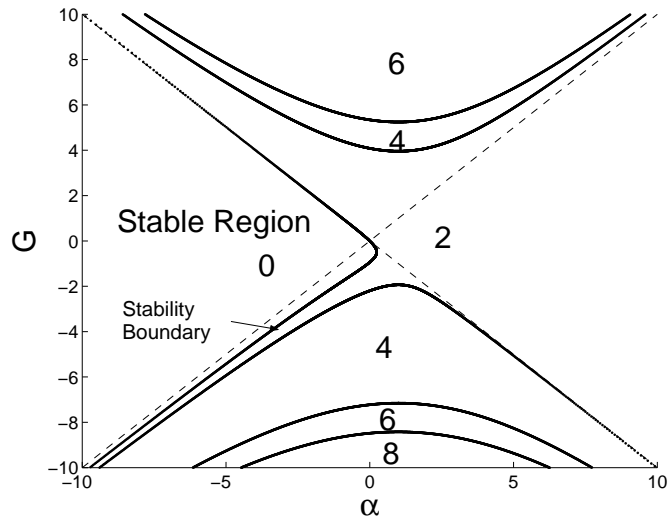}}}
\caption{Example of critical curves (solid lines) in the
  $(\alpha,G)$--parameter plane of~(\ref{eq:Redeq}), given
  by~(\ref{eq:char}), 
in the nondegenerate case $\beta\Delta t\ne n\pi$ for any integer $n$. 
(Here we used $\Delta t=1$ and $\beta=-4\pi/5$.)
The dashed lines are the asymptotes $G=\pm
  \alpha$. The numbers in the various regions indicate how
  many solutions of the characteristic equation~(\ref{eq:char}) have
  positive real part. The stable region is to the left of all
  critical curves~\cite{RSJ2000}. }
\label{fig:reddylike}
\end{figure}

\begin{figure}
\centerline{\resizebox{2.5 in}{!}{\includegraphics{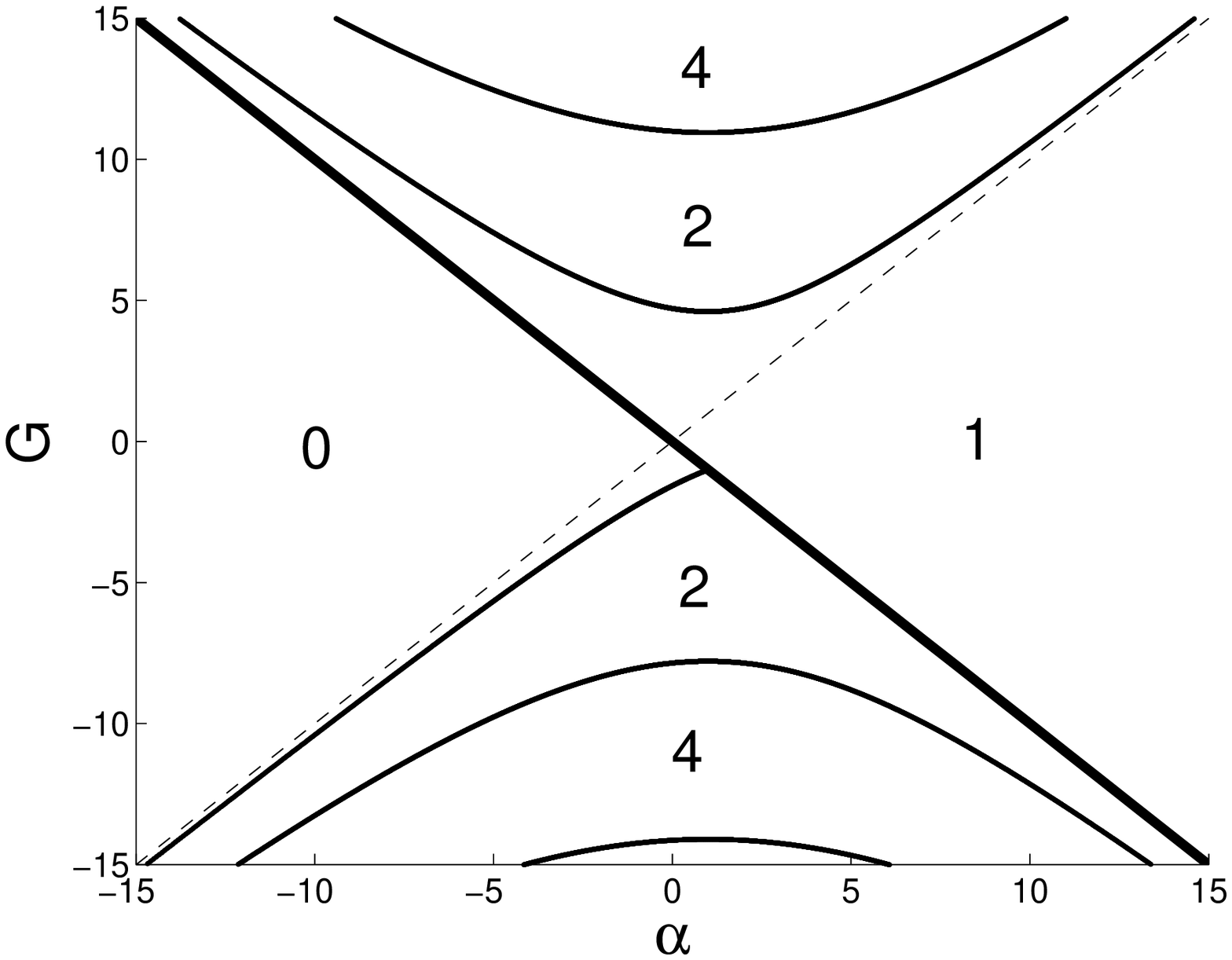}}
\resizebox{2.5 in}{!}{\includegraphics{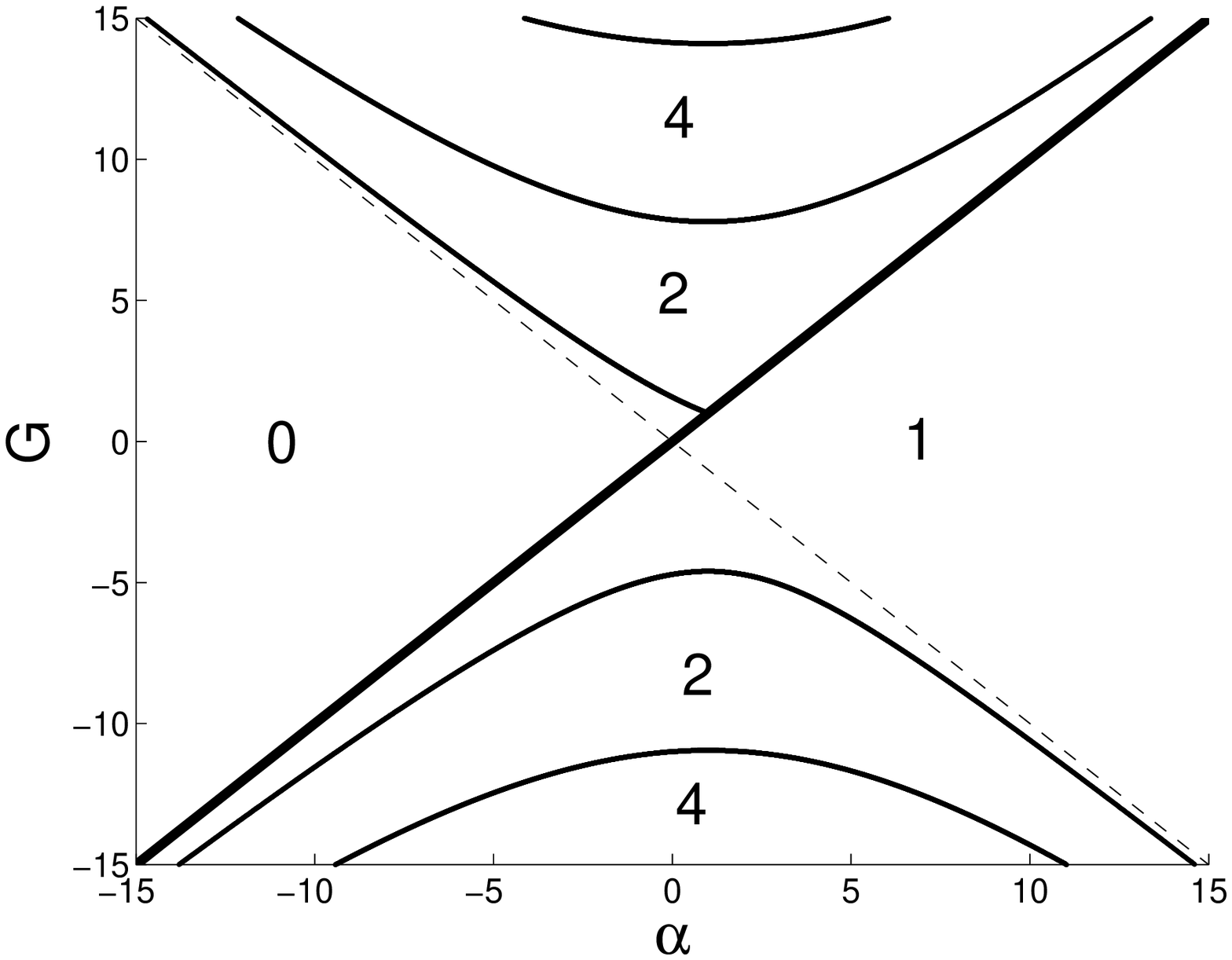}}}
\centerline{$\quad$ (a) \hspace{2.2in} (b)}
\caption{Examples of critical curves associated with~(\ref{eq:Redeq})  
in the degenerate cases for which $\beta\Delta t= n\pi$ 
for some integer $n$. 
(a) $\Delta t=1$, $\beta=-\pi$ (representative of $n$ odd). 
(b) $\Delta t=1$, $\beta=- 2 \pi$ (representative of $n$ even).
}
\label{fig:degenerate}
\end{figure}
Our linear stability analysis builds on some results of Reddy~{\it et
  al.}~\cite{RSJ2000}, who considered the effects of a time--delayed
linear feedback term added to a Hopf bifurcation normal form.  Of
relevance to our analysis are their results on the linear stability of
the origin, which is determined by an analysis of
\begin{equation}
\dot{z}(t) =  (\alpha +i \beta)z(t) -G z(t-\Delta t)\ , \label{eq:Redeq}
\end{equation} 
where $\alpha,\ \beta,\ G,\ \Delta t$ are real parameters with 
$\Delta t>0$. The characteristic equation associated with solutions
of the form $z=e^{\lambda t/\Delta t}$ is
\begin{equation}
\lambda=(\alpha+i\beta)\Delta t-G\Delta te^{-\lambda}\ .
\label{eq:char}
\end{equation}
There are an infinite number of solutions $\lambda$ to this
transcendental equation; if all of them satisfy Re($\lambda$)$<$0,
then $z=0$ is stable. If any solution $\lambda$ has positive real
part, then $z=0$ is unstable.  (See, for example, Diekmann~{\it et
al.}~\cite{Diek} or Driver~\cite{Driver} for background on delay
equations.)

A solution $\lambda$ of the characteristic equation is purely
imaginary on the ``critical curves'' in the $(\alpha,G)$--parameter
plane. These curves are significant because they represent boundaries
in the $(\alpha,G)$--plane across which the number of solutions with
Re$(\lambda)>0$ changes.  Figure~\ref{fig:reddylike} shows an example
of critical curves in the $(\alpha,G)$--plane, associated with a
``nondegenerate case'' for which $\beta\Delta t \ne n\pi$ for any
$n\in Z$, with examples of critical curves for degenerate cases
provided in Figure~\ref{fig:degenerate}.  A parameterization of the
critical curves is obtained by letting $\lambda=i\nu$, $\nu\in R$, in
(\ref{eq:char}), in which case we find
\begin{eqnarray}
\alpha&=&{(\nu-\beta\Delta t)\cos(\nu)\over\Delta t\sin(\nu)},\nonumber\\ 
G&=& {(\nu-\beta\Delta t)\over\Delta t \sin(\nu)}\ .
\label{eq:crit}
\end{eqnarray}  
Here each critical curve is associated with an interval
$\nu\in(m\pi,(m+1)\pi)$ for some integer $m$.  In the nondegenerate
situation, Reddy~{\it et al.}~\cite{RSJ2000}, extending a result in
\cite{Diek}, show that the stability boundary is determined by the
left most critical curve in the $(\alpha,G)$--plane, as indicated in
Figure~\ref{fig:reddylike}.  This stability boundary corresponds to
the critical curve for the $\nu$--interval containing $\beta\Delta t$;
it passes through the origin and approaches the asymptotes $G=\pm
\alpha$ as $\alpha \rightarrow -\infty$.  In the degenerate cases
(Figure~\ref{fig:degenerate}), defined by $\beta\Delta t=n\pi$ for
some integer $n$, there is an additional critical curve coincident
with one of the asymptotes; specifically if $n$ is even (odd) then on
the critical curve $\alpha=G$ ($\alpha=-G$) the characteristic
equation~(\ref{eq:char}) is solved by $\lambda=i\nu=i\beta\Delta
t(=in\pi$). Finally, we show that the following properties
hold for the  critical curves in the nondegenerate case.

\begin{lemma}
\label{tm:othercurves}
Consider $G<0$ and $\beta\Delta t\ne n\pi $ for any $n\in Z$
in~(\ref{eq:Redeq}).  (1) The stability boundary in the
$(\alpha,G)$--parameter plane is the greatest distance from the
asymptote $\alpha=G$ for $G=-\frac{1}{\Delta t}$.  (2) All other
critical curves (to the right of the stability boundary and for $G<0$) 
lie in the region $G<-\frac{1}{\Delta t}$.
\end{lemma}

\begin{proof}
(1) At a point where the distance between the stability boundary and
the asymptote $\alpha=G$ is maximized
\begin{equation}
\frac{d\alpha}{d\nu}=\frac{dG}{d\nu}\ ;\quad
\frac{d^2G}{d\nu^2}-\frac{d^2\alpha}{d\nu^2}>0\ ,
\end{equation}
where $\alpha(\nu)$ and $G(\nu)$ are the parametric equations for the
critical curves~(\ref{eq:crit}). It follows by direct calculation
that these conditions are met when $\nu=\nu^*$, where $\nu^*$ is defined
implicitly by the equation
\begin{equation}
\sin(\nu^*)=\beta\Delta t-\nu^*\ .
\end{equation}
{F}rom~(\ref{eq:crit}), we see that this implies $G=-\frac{1}{\Delta
  t}$. (It follows from property (2) of this lemma that $\nu^*$
defines a point on the stability boundary rather than another critical
curve.)  (2) The remaining critical curves in the region  $G<0$
reach their maximum in the $(\alpha,G)$-plane at points for which
$\frac{dG}{d \nu} = 0$ (while $\frac{d\alpha}{d\nu}\ne 0$), where
\begin{equation}
   \frac{dG}{d \nu} = \frac{\sin(\nu) - (\nu-\beta\Delta t)\cos(\nu)}
{\Delta t\sin^2(\nu)}\ .
\end{equation}
Substituting
\begin{equation}
\sin(\nu) = (\nu-\beta\Delta t) \cos(\nu)
\label{eq:numax}
\end{equation}
into~(\ref{eq:crit}) yields
\begin{equation}
   G_{max}=\frac{1}{\Delta t \cos(\nu)}
\end{equation}
for the maximum value of $G$, where $\nu$ satisfies~(\ref{eq:numax}). 
For the critical curves confined to $G<0$, $-1<\cos(\nu)<0$ in~(\ref{eq:numax})
so $G_{max}<-\frac{1}{\Delta t}$.
\qquad\end{proof}

\section{Linear Stability Analysis}
\label{sec:analysis}

\subsection{Preliminaries}
\label{sec:prelims}

In order to determine whether the
feedback~(\ref{eq:spatialfb}-\ref{eq:2pi}) can stabilize a traveling
wave solution~(\ref{eq:ATW}) of~(\ref{eq:mainamp}), we analyze the
effects of small amplitude perturbations on the solution by letting
\begin{equation}
A = R e^{iKx+i \omega t} ( 1 + a_{+}(t) e^{iQx} +a_{-}(t)e^{-iQx}). 
\label{eq:anz}
\end{equation}
Here $a_+$ and $a_-$ are the amplitudes of 
small perturbations with wavenumbers $K \pm Q$. (We may,  without loss of
generality, assume that $Q\ge 0$.)  
Substituting ($\ref{eq:anz}$) into ($\ref{eq:mainamp}$) and linearizing, 
yields the following system of delay equations:
\begin{equation}  \frac{d}{dt} \left( \begin{array}{l}
   a_{+}(t) \\
   a_{-}^{*}(t) 
   \end{array} \right) = J
   \left(\begin{array}{l}
   a_{+}(t) \\
   a_{-}^{*}(t)  
   \end{array}\right) - \gamma 
  \left(\begin{array}{l}
   a_{+}(t- \Delta t) \\
   a_{-}^{*}(t- \Delta t)  
   \end{array}\right)\ ,             \label{eq:timesystem}
\end{equation}
where
\begin{equation}
 J = \left[ \begin{array}{cc}
    \gamma-Q(2K+Q)(1+ib_{1})   &  -(b_{3}-i)R^{2} \\ 
    -(b_{3}-i)R^{2}+\rho [e^{iQ\Delta x} -1]& \\
    & \\
    -(b_{3}+i)R^{2} & \gamma+Q(2K-Q)(1-ib_{1})  \\
    & -(b_{3}+i)R^{2}+\rho [e^{iQ\Delta x} -1]
\end{array}\right] \ . \label{eq:Jmatrix} \end{equation}
Diagonalizing yields two decoupled linear delay equations,
\begin{equation}
  \frac{d}{dt} c_{k}(t) = j_{k} c_{k}(t) - \gamma c_{k}(t - \Delta t)\ , 
\quad k=1,2, 
\label{eq:deleq1}\end{equation}
where $j_1$ and $j_2$ are the complex eigenvalues of $J$ defined such
that $Re(j_1) \geq Re(j_2)$. 

Note that the temporal and spatial feedback terms, proportional to
$\gamma$ and $\rho$, simply shift the eigenvalues of $J$ by the
diagonal factor $\gamma+\rho[e^{iQ\Delta x}-1]$. In particular, each
$j_k$ is of the form $j_k=\gamma+\widehat{j_k}$, where $\widehat{j_k}$
is independent of $\gamma$ and depends on $b_1,\ b_3,\ \rho$, $K$, and
$Q$. In the following we denote the real and imaginary parts of
$\widehat{j_k}$ by $\widehat{j_{kr}}$ and $\widehat{j_{ki}}$,
respectively.  Since $\widehat{j_{1r}}$ plays a central role in our
analysis, we simplify notation by replacing it with $f$ and
suppressing its dependence on the system and solution parameters
$b_1,b_3,\rho,K$, viewing it as a function of the perturbation
parameter $Q$. Specifically, we let
\begin{equation}
f(Q)\equiv \widehat{j_{1r}}(Q;b_1,b_3,\rho,K)\equiv 
j_{1r}(Q;b_1,b_3,\rho,K,\gamma)
-\gamma\ .
\label{eq:fqdef}
\end{equation}
We note that $f(Q)$ and $j_{1i}(Q)$ are continuous functions of $Q$, and
that $f(Q)$ satisfies
\begin{eqnarray}
\qquad f(0)&=&0,\nonumber\\
\lim_{Q \rightarrow \infty} f(Q) &\rightarrow& -\infty\ .
\end{eqnarray}
Hence $f(Q)$ has an absolute maximum $f_{max}\ge 0$ for some $Q\ge 0$.
The significance of $f(Q)$ is that it represents the growth rate of
perturbations of wavenumber $Q$ in the absence of the time-delay feedback
({\it i.e.} for $\gamma=0$).

The characteristic equations associated with solutions
$c_k(t)=e^{\lambda_k t/\Delta t}$ of~(\ref{eq:deleq1}) are
\begin{equation}
\lambda_k=\Delta t(j_k -\gamma e^{-\lambda_k}),\quad k=1,2\ .
\label{eq:char2}\end{equation}
We obtain the critical curves associated with our linear stability
problem~(\ref{eq:deleq1}) by substituting $\lambda_k=i \nu_k$
into~(\ref{eq:char2}). Then, for instance, we find the following
parametric representation of critical curves in the
$(j_{1r},\gamma)$--planes:
\begin{equation}
j_{1r} = \frac{(\nu_1 - j_{1i} \Delta t) \cos(\nu_1)}{ \Delta t \sin( \nu_1 )} 
\label{eq:j1rpar}
\end{equation}
\begin{equation}
\gamma = \frac{(\nu_1 - j_{1i} \Delta t)}{ \Delta t \sin( \nu_1 )}. 
\label{eq:g1par}
\end{equation}
Similar equations apply for the critical curves in the
$(j_{2r},\gamma)$--parameter plane. Using $j_{1r}=\gamma+f(Q)$ (which,
recall, defines $f(Q)$) we determine that on the critical curves in
the $(j_{1r},\gamma)$--plane $Q$ and $\nu_1$ are related through
\begin{equation}
f(Q) \Delta t = w(\nu_1,Q),  
\label{eq:QV1}
\end{equation}
where
\begin{equation}
w(\nu_1,Q)\equiv
\frac{(\nu_1 - j_{1i}(Q) \Delta t) 
(\cos(\nu_1)-1)}{\sin(\nu_1)}\ ,
\label{eq:QV2}
\end{equation}
which is a result we use later in our analysis.

\begin{figure}
  \centerline{\resizebox{2.55 in}{!}{\includegraphics{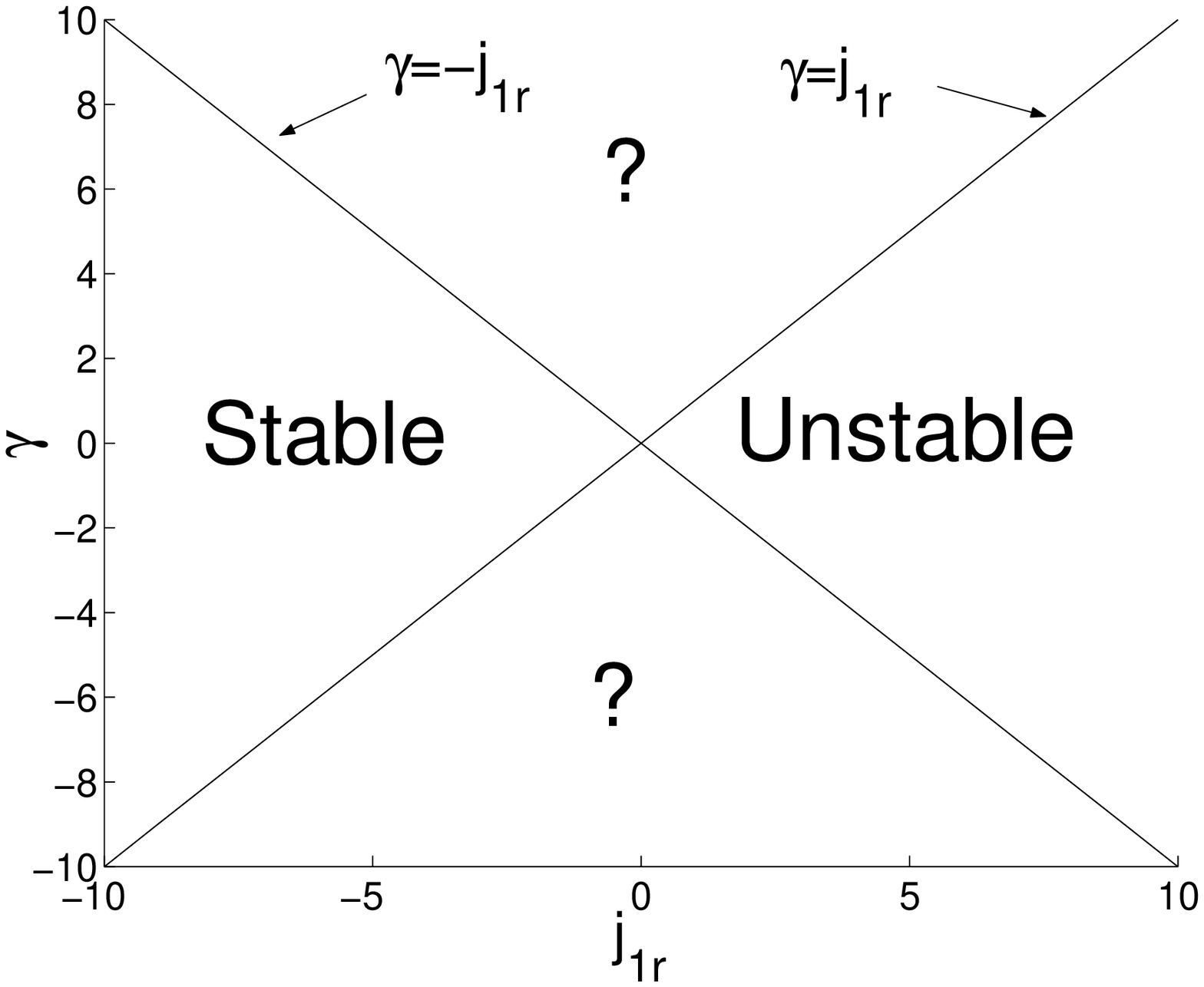}}
\resizebox{2.35 in}{!}{\includegraphics{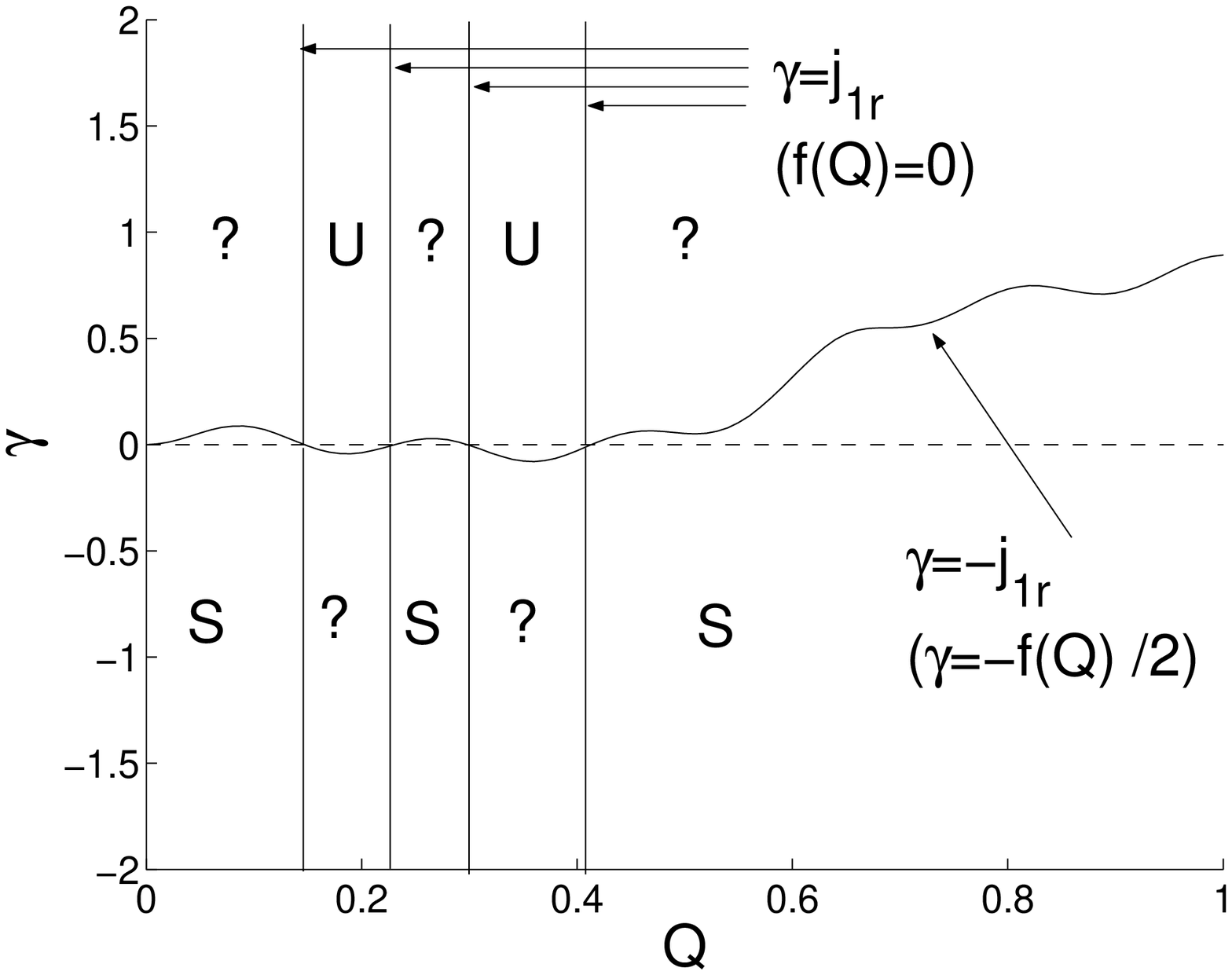}}}
\centerline{$\quad$ (a) \hspace{2.2in} (b)}
\caption{Mapping of the guaranteed stable (S) 
and unstable (U) regions bounded by $\gamma=\pm j_{1r}$ from 
(a) the $(j_{1r},\gamma)$--plane to (b) the $(Q,\gamma)$--plane.
The parameter values are
$b_1=5$, $b_3=1.2$, $K =0.18$, and $\rho=0.1$.}
   \label{fig:comparespace}  
\end{figure}

The parametric equations for the critical
curves~(\ref{eq:j1rpar}-\ref{eq:g1par}) in the
$(j_{1r},\gamma)$--plane are similar to those obtained by Reddy~{\it
  et al.}~\cite{RSJ2000}, given by~(\ref{eq:crit}). However since both
$j_{1r}$ and $j_{1i}$ depend on the perturbation wavenumber $Q$ the
results in~\cite{RSJ2000}, summarized by Figures~\ref{fig:reddylike}
and~\ref{fig:degenerate}, do not apply directly. Nonetheless some
useful results concerning the stability boundary do carry over. For
instance, it follows from~(\ref{eq:j1rpar}) and~(\ref{eq:g1par}) that
$j_{1r} = \gamma \cos(\nu)$ so all critical curves must lie in the
region of the ($j_{1r}$,$\gamma$)--plane where $|j_{1r}| \leq
|\gamma|$. In other words, if we divide the ($j_{1r},\gamma$)--plane
into four quadrants separated by $\gamma = \pm j_{1r}$ the critical
curves lie inside (or possibly on the boundaries) of the upper and
lower quadrants (as is also the case in Figures~\ref{fig:reddylike}
and \ref{fig:degenerate}).  The stability assignments are necessarily
the same at all points inside the left and right quadrants of the
parameter plane.  In particular, the left quadrant is contained in the
stable region and the right quadrant is contained in the unstable
region for all $j_{1i}$. Because the stability boundary is contained
in the upper and lower quadrants both stable and unstable regions are
located there, as signified by the question marks in
Figure~\ref{fig:comparespace}. For fixed values of $b_1$, $b_3$, $K$,
and $\rho$, the quadrants of the ($j_{1r}$,$\gamma$)--plane may be
mapped to the ($Q$,$\gamma$)--plane.  In particular, the asymptotes
$\gamma = \pm j_{1r}$ are mapped to the curves defined by $f(Q)=0$ and
$\gamma=-f(Q)/2$; an example is shown in
Figure~\ref{fig:comparespace}.

We note that due to the translation symmetry of~(\ref{eq:mainamp}),
which acts nontrivially on the traveling wave solution~(\ref{eq:ATW}),
there is always a neutral mode associated with perturbations at
$Q=0$. (It is this symmetry that forces $f(0)=0$.)  Thus the line
$j_{1r}=\gamma$ is a (degenerate) critical curve; as it is a
consequence of translation symmetry, it does not necessarily herald an
instability. 

We now make use of the elementary (in)stability results summarized by
Fig.~\ref{fig:comparespace}(a) to prove
Lemmas~\ref{lemma-fpos}--\ref{tm:ap1}, which allow us to restrict any
further linear stability analysis to the $k=1$ delay equation
of~(\ref{eq:deleq1}) for $Q>0$, and for feedback parameters $\gamma<0$
and $\rho\ge 0$. In particular, we find that the restriction
$\gamma<0$, $\rho>0$ is sufficient to ensure that the feedback does
not further destabilize the traveling waves in the Benjamin-Feir
unstable regime, {\it i.e.} the feedback does not introduce new
instabilities in this case.

\begin{lemma}
\label{lemma-fpos}
(a) If $f(Q)<0$ for all $Q>0$, where $f(Q)$ is defined by the
relation~(\ref{eq:fqdef}), then the traveling wave is stabilized for
any $\gamma\le 0$.  In particular, the wave is stabilized using
spatial feedback alone in~(\ref{eq:spatialfb}) ({\it i.e.} for
$\gamma=0$), and there are no instabilities associated with $Q=0$ when
$\gamma\le 0$. (b) If $f(Q)>0$ for any $Q$ then the traveling wave
cannot be stabilized using $\gamma\ge 0$.
\end{lemma}

\begin{proof}
  (a) If $f(Q)<0$, then $j_{2r}\le j_{1r}<\gamma$ for all $Q>0$ since
  $j_{1r}=\gamma+f(Q)$. Hence, for $\gamma\le 0$, the system is in the
  guaranteed stable regime of Fig.~\ref{fig:comparespace}(a) for all
  perturbations with $Q>0$. It remains to eliminate the possibility of
  instabilities at $Q=0$, where $j_{2r}(0)<j_{1r}(0)=\gamma$. For this
  we consider the solutions $\lambda_k=\mu_k+i\nu_k$ of the
  characteristic equation~(\ref{eq:char2}), and determine
  that the growth rates $\mu_k$ satisfy
\begin{equation}
\mu_k=\Delta t\Bigl(\widehat{j_{kr}}+\gamma(1- e^{-\mu_k}\cos(\nu_k))\Bigr)\ .
\label{eq:growths}
\end{equation}
There are no $\mu_k>0$ solutions of~(\ref{eq:growths}) when $\gamma\le
0$ and $\widehat{j_{kr}}\le 0$. This is proved by assuming there is a
solution with $\mu_k> 0$, which would imply that the left-- and
right--hand--sides of~(\ref{eq:growths}) do not have the same sign,
thereby leading to a contradiction. In fact, the only solution with
$\mu_k=0$ is the one forced by translation symmetry. Hence there can
be no linear instabilities associated with $Q=0$ when $\gamma\le 0$.

(b) If $\gamma>0$ and $f(Q)>0$ for some $Q$, then the system is in the
guaranteed unstable regime of Fig.~\ref{fig:comparespace}, defined
by $\gamma>-{1\over 2}f(Q)$, $f(Q)>0$, for perturbations with that
wavenumber.  If there is no temporal feedback ($\gamma=0$) and
$f(Q)>0$ for some $Q$, then instability for these $Q$ values follows
directly from the $c_1(t)$ equation of~(\ref{eq:deleq1}), which is
\begin{equation}
\frac{dc_{1}}{dt} = (f(Q)+ij_{1i}) c_{1}\quad {\rm for} \quad \gamma=0\ .
\label{eq:c1gam0}
\end{equation}\qquad\end{proof}

Lemma~\ref{lemma-fpos} motivates our focus on $\gamma<0$ in the
remainder of the paper: if the time--delay feedback is to be effective
in stabilizing plane waves, then the value of $\gamma$ used
in~(\ref{eq:spatialfb}) should be negative. We similarly focus on
$\rho\ge 0$. This is because the spatial feedback has the effect of
shifting each $j_{kr}$ in~(\ref{eq:growths}) by $\rho[\cos(Q\Delta
x)-1]$.  Thus there is no effect of the spatial feedback on
perturbations with wavenumber $Q=2 n\pi/\Delta x$ for some $n\in Z$.
And, for all other wavenumbers, there is a shift towards the stable
region of Figure~\ref{fig:comparespace}(a) if $\rho>0$ and towards the
unstable region if $\rho<0$. In particular, in the absence of
time--delay feedback, if $\rho<0$ the spatial feedback exacerbates
longwave instabilities in the Benjamin--Feir unstable regime. In the
remainder of the paper we consider only $\rho\ge 0$.  In this case, as
explained by the following lemma, we may ignore the $c_2$--equation
of~(\ref{eq:deleq1}) in the remainder of our linear stability
analysis.

\begin{lemma}
\label{tm:ap1}
If $\gamma\le 0$ and $\rho\ge 0$ 
in~(\ref{eq:deleq1}) then all solutions of the $c_2$--equation
of~(\ref{eq:deleq1}) decay exponentially.
\end{lemma}

\begin{proof}
The real parts of the eigenvalues $j_k$ of the matrix
$J$~(\ref{eq:Jmatrix}),
can be written in the form 
\begin{equation}
  j_{1r} = a +\gamma + |p|\ ,\quad 
  j_{2r} = a +\gamma - |p|\ 
\end{equation}
since they are solutions of a quadratic. From~(\ref{eq:Jmatrix}) it
follows that
\begin{equation}
  a = -\Bigl(Q^2 +b_3 R^2 +\rho[1-\cos(Q\Delta x)]\Bigr),\ 
\end{equation}
so $a<0$ for $\rho\ge 0$.
Thus  
\begin{equation}
  j_{2r}-\gamma = a - |p| <0
  \label{eq:j2cond}
\end{equation}
for all values of $Q$ and $\rho\ge 0$. 
{F}rom Fig.~\ref{fig:comparespace}(a) (with $j_{2r}$
replacing $j_{1r}$) we see that the parameters are in the guaranteed
stable region when $\gamma<0$.  
Finally, if $\gamma=0$, then the $c_2$--equation
reduces to 
\begin{equation}
\frac{d}{dt} c_{2} = \widehat{j_{2}} c_{2}\quad {\rm for} \quad \gamma=0\ ,
\label{eq:c2gam0}
\end{equation}
so all solutions decay for $\gamma=0$ since 
$\widehat{j_{2r}}=a-|p|<0$.
 \qquad
\end{proof}

\subsection{Main Stability Results}
\label{sec:NS}

Lemma~\ref{lemma-fpos} states that if $f(Q)$, defined
by~(\ref{eq:fqdef}), is negative for all $Q>0$, then the traveling
wave may be stabilized using spatial feedback alone ($\gamma=0$).
This section addresses the more difficult case where the chosen
spatial feedback fails to stabilize the traveling wave, {\it i.e.}, we
consider the situation where $f(Q)>0$ for some range of $Q$ values and
investigate whether the traveling waves may then be stabilized by the
addition of an appropriate time--delay feedback with $\gamma<0$.
({F}rom Lemma~\ref{lemma-fpos} we know that the wave cannot be
stabilized with $\gamma>0$.)  We focus on the case where the spatial
feedback parameter $\rho$ is fixed at some non--negative value so that
Lemma~\ref{tm:ap1} applies. We show that it is then sufficient to
consider just $\gamma=-1/\Delta t$ in order to determine whether or
not the traveling waves can be stabilized by including the time--delay
feedback in~(\ref{eq:spatialfb}).

\begin{figure} 
  \centerline{\resizebox{2.6in}{!}{\parbox{2.6in}{}
\includegraphics{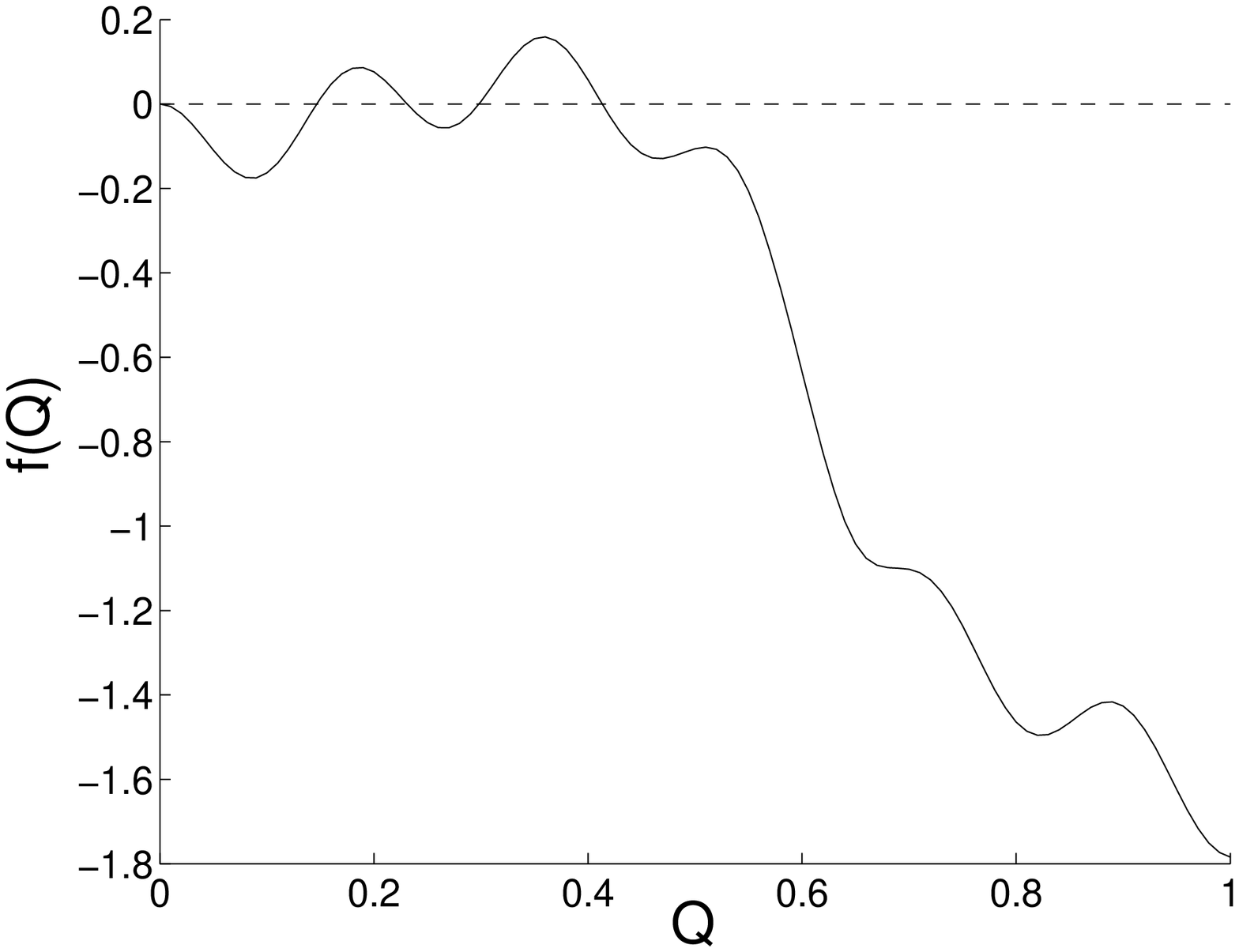}}
\resizebox{2.6in}{!}{\includegraphics{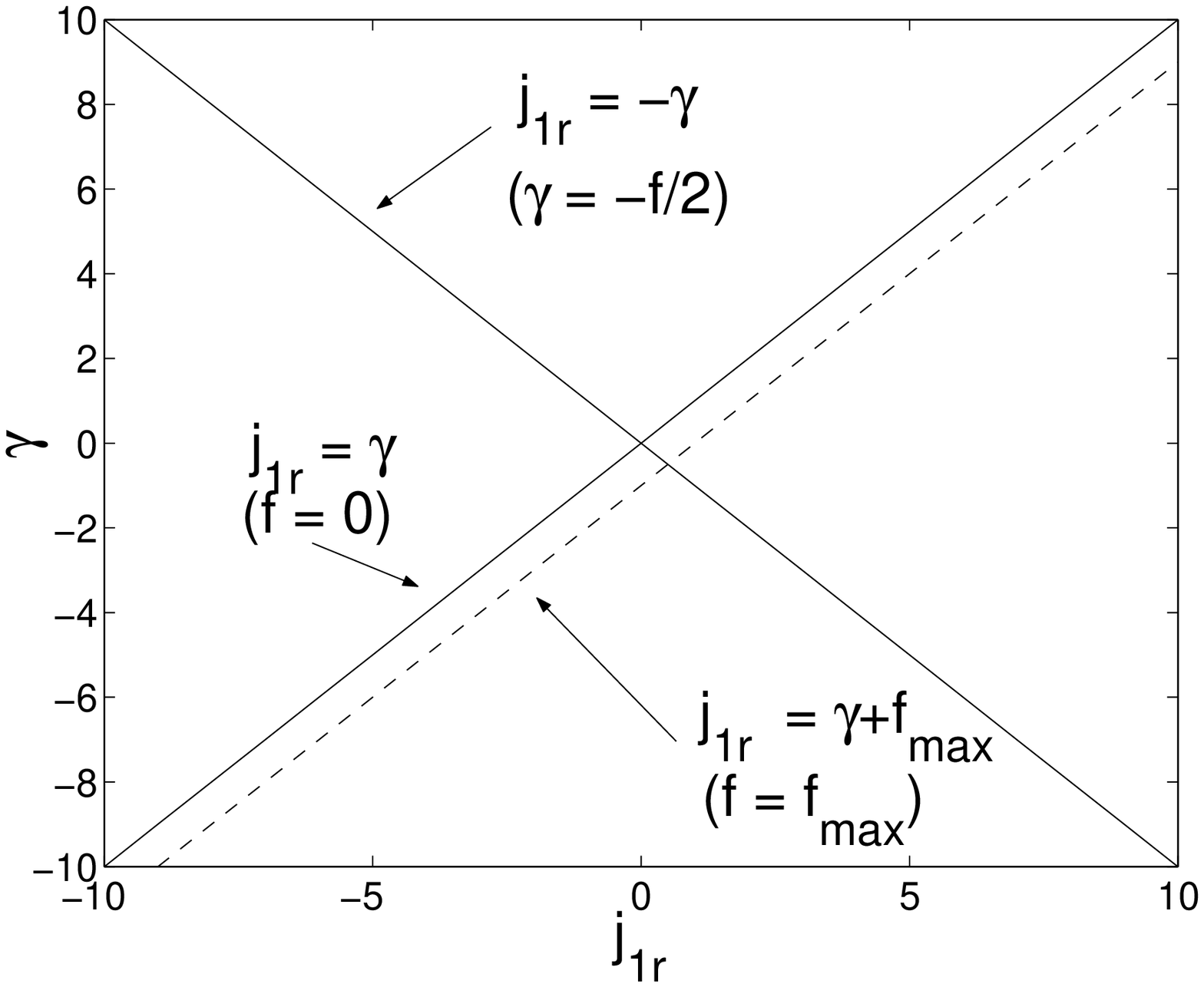}}}
\centerline{$\quad$ (a) \hspace{2.2in} (b)}
 \caption{(a) Plot of $f(Q)$ for the parameters 
of Figure~\ref{fig:comparespace}. 
(b) Corresponding regions of the $(j_{1r},\gamma)$--parameter plane 
(where $j_{1r}=\gamma+f(Q)$); 
$j_{1r}$ values to the right of the existence 
line~(\ref{eq:existence-boundary}) are inaccessible. 
}
\label{fig:exist}
\end{figure}

\begin{figure}
\centerline{\resizebox{2.6 in}{!}{\includegraphics{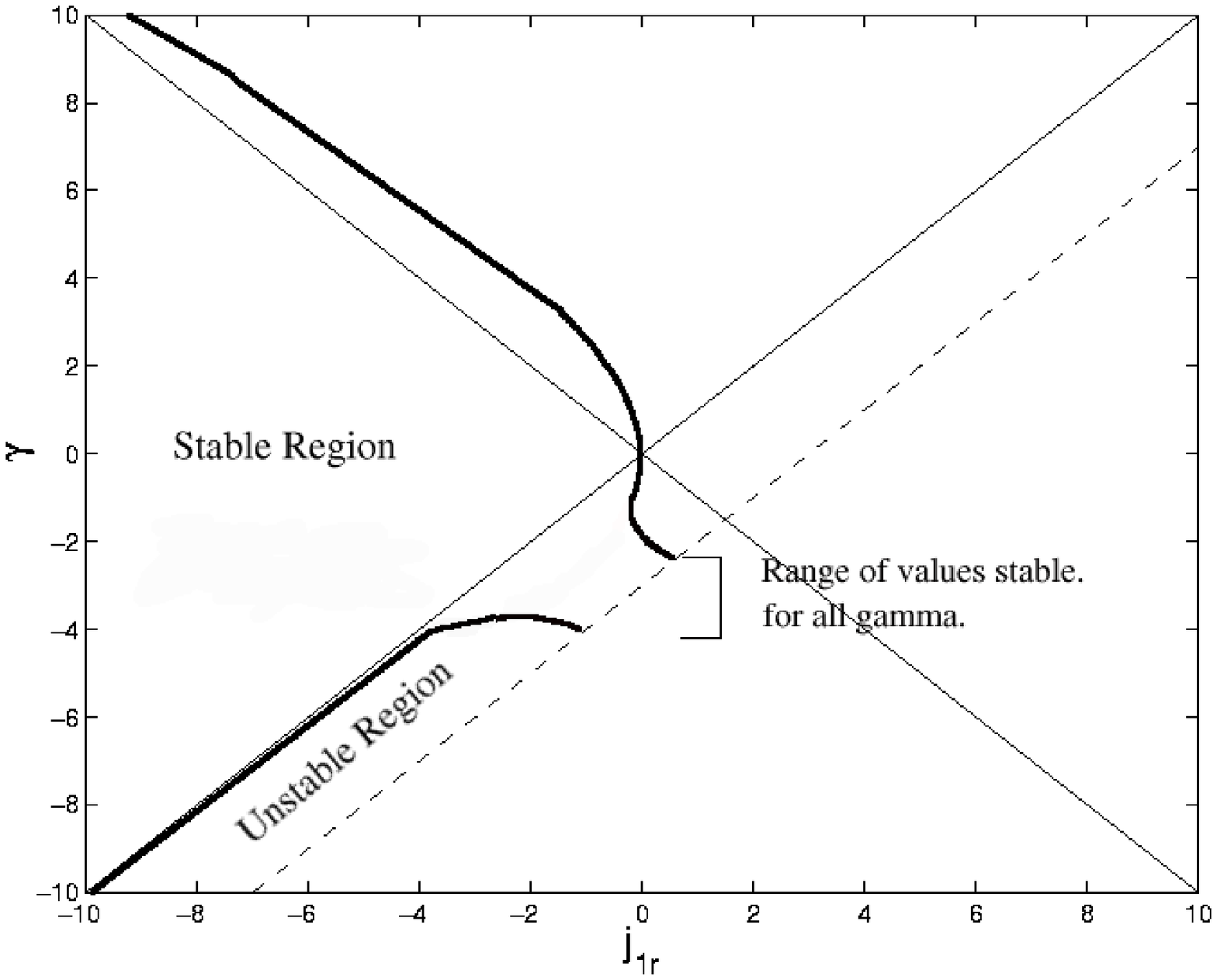}}
\resizebox{2.6 in}{!}{\includegraphics{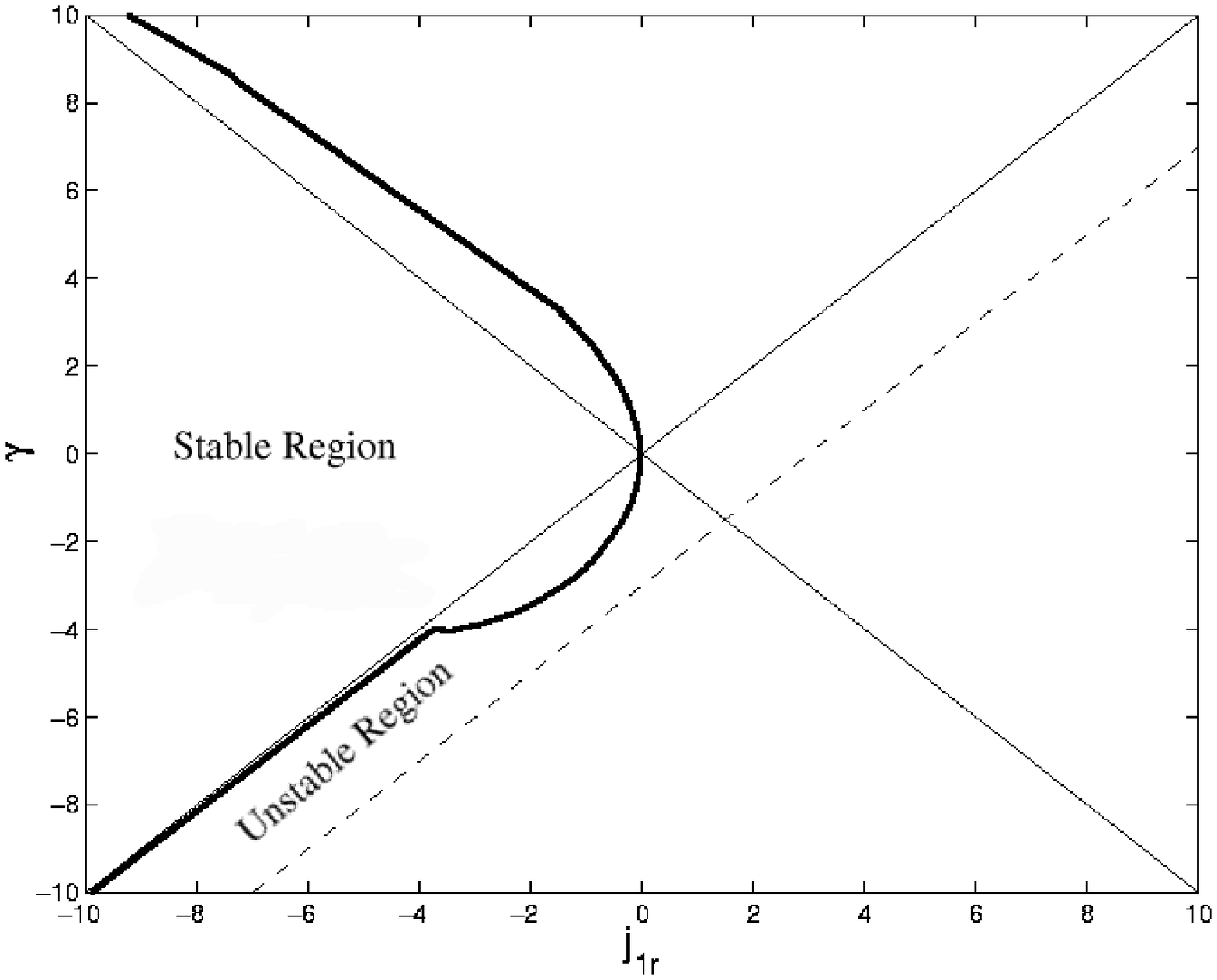}}
}
\centerline{$\quad$ (a) \hspace{2.2in} (b)}
\caption{(a) Schematic of stable region for case that 
the stability boundary leaves 
  the existence region. Range of $\gamma$ values for which
the traveling wave is stable against
  perturbations of all wavenumber $Q$ is indicated.
(b) Schematic of the unstable case where  the stability
  boundary lies to the left of the existence line for all $\gamma$.
}
\label{fig:outbound}
\end{figure}

Our analysis relies on the observation that not all regions of the
$(j_{1r},\gamma)$--parameter plane are accessible for fixed
$b_1,b_3,K,\rho$ and $Q\ge 0$. In particular, since
$j_{1r}=\gamma+f(Q)$ and $f(Q)$ has an absolute maximum (denoted by
$f_{max}$), parameter values to the right of the line 
\begin{equation}
\gamma=j_{1r} - f_{max}
\label{eq:existence-boundary}
\end{equation} 
are never achieved for any value of $Q$ (see Figure \ref{fig:exist}).
We refer to the boundary~(\ref{eq:existence-boundary}) between
accessible and inaccessible regions of the
$(j_{1r},\gamma)$--parameter plane as the ``existence line'', meaning
that there {\it exist} values of $Q$ that place $j_{1r}$ and $\gamma$
on or to the left of this line, but never to the right of it. Our
stability analysis focuses on determining whether there is a
(negative) $\gamma$--value such that the traveling wave is stable for
all values of $j_{1r}$ between $-\infty$ and the existence line, {\it
  i.e.} for all values of $Q$.  Fig.~\ref{fig:outbound} presents,
schematically, projections onto the $(j_{1r},\gamma)$--plane of the
points that lie on the stability boundary in the three-dimensional
$(j_{1r}(Q),\gamma,j_{1i}(Q))$ parameter space.
Figure~\ref{fig:outbound}(a) typifies the case where it
is possible to stabilize the traveling wave over the range of $\gamma$--values
for which the stability boundary leaves the existence region.
On the other hand,  in Fig.~\ref{fig:outbound}(b) 
the stability boundary lies entirely to the left of the existence line
for all values of $\gamma$ so it is not possible to stabilize the
traveling wave in this case.

We now state our main linear stability result; the proof, which relies on
Lemma~\ref{tm:unstab} below, 
is deferred to the end of this section.

\begin{theorem}
\label{tm:characteristic}
The traveling wave solution~(\ref{eq:ATW}) of the CGLE~(\ref{eq:CGLE})
with feedback~(\ref{eq:spatialfb}-\ref{eq:2pi}) may be (linearly)
stabilized, for a fixed given value of $\rho\ge 0$, if and only if 
the following equations {\it fail} to
have a real solution $(\nu,Q)$, $Q>0$:
\begin{equation}
\nu+\sin(\nu)=\Delta t\ j_{1i}(Q) 
\label{eq:crit1}
\end{equation}
\begin{equation}                   
1-\cos(\nu)=\Delta t\ f(Q)\ .
\label{eq:crit2}
\end{equation}
If a solution fails to exist then the traveling wave can be stabilized
using $\gamma=-1/\Delta t$.
\end{theorem}

{\bf Remark 1}: Note that any solution of~(\ref{eq:crit2}) that exists
must lie in a closed interval of $Q$ values defined by the requirement
$f(Q)\Delta t\in [0,2]$.

{\bf Remark 2}: Equations~(\ref{eq:crit1}--\ref{eq:crit2}) are
equivalent to the equations for the critical
curves~(\ref{eq:j1rpar}-\ref{eq:g1par}) for $\gamma=-1/\Delta t$ and
$\nu_1=\nu$.

The following lemma addresses special degenerate situations for which
stabilization of the traveling wave is impossible.

\begin{lemma}
\label{tm:degeneracies}
Let $Q=Q_{2n}>0$ be defined by the condition that 
$j_{1i}(Q_{2n})\Delta t=2\pi n$,
$n\in Z$.  The traveling wave solution~(\ref{eq:ATW}) of the
CGLE~(\ref{eq:CGLE}) with feedback~(\ref{eq:spatialfb}-\ref{eq:2pi})
is linearly unstable for all values of $\gamma$ if for any $n$ (a)
$f(Q_{2n})>0$, or (b) $f(Q_{2n})=0$ and $\frac{df}{dQ}(Q_{2n})\ne 0$.  (c) If
$f(Q_{2n})=\frac{df}{dQ}(Q_{2n})=0$, then the traveling wave is either
linearly unstable or neutrally stable; it is not linearly stable for
any value of $\gamma$.
\end{lemma}

\begin{proof}
(a) In the degenerate situation, for the value of $Q=Q_{2n}$, where
 $j_{1i}(Q_{2n})\Delta t=2\pi n$ ($n\in Z$), the line $j_{1r}=\gamma$
 represents a critical curve for the delay equation~(\ref{eq:deleq1})
 (with $k=1$). (Here we are viewing $j_{1r}$ as an independent
 parameter, with $j_{1i}$ held fixed; see Fig.~\ref{fig:degenerate}(b)
 with $\alpha=j_{1r}$ and $G=\gamma$.) Thus the traveling wave
 solution is unstable to perturbations with wavenumber $Q=Q_{2n}$ if
 $j_{1r}(Q_{2n})>\gamma$. Since $f(Q)\equiv j_{1r}(Q)-\gamma$, this
 inequality is met whenever $f(Q_{2n})>0$.

(b) If $f(Q_{2n})=0$, then $j_{1r}(Q_{2n})=\gamma$ and the solution is
neutrally stable to perturbations of wavenumber $Q=Q_{2n}$ for
$\gamma\le 0$. (It is unstable if $\gamma>0$ since it lies in the guaranteed
unstable regime of Fig.~\ref{fig:comparespace}(a)) To determine the stability
of the traveling wave for $\gamma< 0$ we examine the movement, with
$Q$, of the solutions $\lambda_1=\mu_1+i\nu_1$ of the
characteristic equation~(\ref{eq:char2}). {F}rom the real and imaginary
parts of~(\ref{eq:char2}) we have
\begin{equation}
  \mu_1 = f(Q)
\Delta t+\gamma \Delta t - \gamma \Delta t e^{-\mu_1} \cos(\nu_1), 
\label{eq:fullchar2real}
\end{equation} 
\begin{equation}
  \nu_1 = j_{1i}(Q) \Delta t + \gamma\Delta t e^{-\mu_1} \sin(\nu_1), 
\label{eq:fullchar2imag}
\end{equation}
which has solution $\mu_1=0$, $\nu_1=j_{1i}(Q_{2n})\Delta t=2\pi n$ for $Q=Q_{2n}$.
To determine the movement of $\mu_1$ with $Q$ near $Q_{2n}$, we compute
$\frac{d\mu_1}{dQ}(Q_{2n})$. We find
\begin{equation}
\frac{d\mu_1}{dQ}(Q_{2n})=\frac{\Delta t}{1-\gamma\Delta t}
\frac{df}{dQ}(Q_{2n})\ne 0\ .
\end{equation}
Since $\frac{\Delta t}{1-\gamma\Delta t}>0$ for $\gamma<0$, the 
eigenvalue $\lambda_1$ moves into the unstable region $\mu_1>0$
with increasing (decreasing) $Q$ if $\frac{df}{dQ}(Q_{2n})>0$ ($<0$).

(c) If $f(Q_{2n})=\frac{df}{dQ}(Q_{2n})=0$, then the traveling wave 
may only be neutrally stable for
$\gamma\le 0$. This latter claim follows since the critical eigenvalue
$\lambda_1(Q_{2n})=2\pi n$ does not cross the imaginary axis with finite
speed as $Q$ is varied near $Q_{2n}$, {\it i.e.}
$\frac{d\mu_1}{dQ}(Q_{2n})=0$. This tangency implies that the traveling
wave cannot be linearly stable in this case -- it is either unstable or
``at best''
neutrally stable.
\qquad\end{proof}

A consequence of this lemma is that the temporal feedback cannot be
used to stabilize homogeneous oscillations. (Note that for $K=0$,
spatial feedback is not applicable and only the temporal feedback is
relevant.) Specifically, for $K=\rho=0$, the matrix $J$, given
by~(\ref{eq:Jmatrix}), has eigenvalues that are purely real for $Q$
sufficiently small and hence we have $j_{1i}(Q)=0$ with $f(Q)>0$ for
sufficiently longwave perturbations associated with the Benjamin-Feir
unstable regime. It follows immediately from
Lemma~\ref{tm:degeneracies} (with $n=0$), that the feedback fails to
stabilize the uniform oscillatory mode associated with $K=0$.  In this
context, we note that Beta~{\it et al.}~\cite{BBMRE03} have given a
possible physical explanation for the failure of time-delayed global
feedback to stabilize a uniform oscillatory mode in the case of
diffusion-induced chemical turbulence. Harrington and
Socolar~\cite{HS01} have a related result that shows that time-delay
feedback alone cannot stabilize traveling wave solutions of the
two-dimensional CGLE.

We may further interpret Lemma~\ref{tm:degeneracies} in light of
analogous results for the spatial feedback. Specifically, we recall
that the effect of the spatial feedback on the eigenvalues of the
matrix $J(Q)$, given by~(\ref{eq:Jmatrix}), is to shift them by the
diagonal entry $\rho(e^{iQ\Delta x}-1)$. Thus if there is an unstable
wavenumber $Q=\widetilde Q_n$ for which $\widetilde Q_n\Delta x=2n\pi$ for
some integer $n$, then the spatial feedback can have no stabilizing
effect on that perturbation wavenumber. In this case, the perturbation
wavenumber $\widetilde Q_n$ is ``resonant'' with the underlying
wavenumber $K$ of the traveling wave ($\widetilde Q_n=nK$), and it does
not feel the influence of the feedback.  Lemma~\ref{tm:degeneracies}
leads to an analogous result when the frequency associated with the
perturbation at $Q_{2n}$, namely $j_{1i}(Q_{2n})$, is in
resonance with the temporal delay $\Delta t$, {\it i.e.}
$j_{1i}(Q_{2n})\Delta t$ happens to also be an even multiple of
$\pi$. In that case, the temporal feedback term also has no effect on
the perturbation, {\it i.e.} it cannot suppress the instability
associated with it. Just~{\it et al.}~\cite{JBORB97} refer to such
situations where the method of Pyragas fails as ``torsion-free''. 
These observations about special (nongeneric)
spatial {\it and} temporal resonant cases are summarized by the
following corollary to Lemma~\ref{tm:degeneracies}.

\begin{corollary}
  If, in the absence of any feedback ($\gamma=\rho=0$), there is an
  unstable wavenumber $Q=\widetilde Q$ for which $\widetilde Q\Delta x=2m \pi$
  and $j_{1i}(\widetilde Q)\Delta t=2n\pi$ for some integers $m$ and $n$,
  then the traveling wave cannot be stabilized using the
  feedback~(\ref{eq:spatialfb}) for any value of $\rho$ or $\gamma$.
\label{tm:maincor}
\end{corollary} 

{\bf Remark:} It follows from~(\ref{eq:2pi}) that the resonance
conditions $\widetilde Q\Delta x=2m\pi$ and $j_{1i}(\widetilde
Q)\Delta t=2n\pi$ are equivalent to the conditions $\widetilde Q=mK$
and $j_{1i}(\widetilde Q)=n\omega$, respectively.

It now remains to consider whether stabilization is possible when
$j_{1i}(Q)\Delta t\ne 2\pi n$ for any $n\in Z$ over all intervals of
$Q$ for which $f(Q)>0$. It is consideration of this situation that
leads to the stabilization criterion of
Theorem~\ref{tm:characteristic}. In particular, we aim to determine a
necessary and sufficient condition for the stability boundary in the
$(j_{1r},\gamma)$--plane to be parameterized by $Q$, such that it
extends {\it continuously} from the origin to $\gamma\to -\infty$
along the asymptote $\gamma=j_{1r}$, as in Fig.~\ref{fig:outbound}(a)
Hence we focus on the mapping~(\ref{eq:QV1}--\ref{eq:QV2}) between
$Q$ and $\nu_1$ along the critical curves.

We first describe some properties of the function $w(\nu_1,Q)$, given
by~(\ref{eq:QV2}), for fixed $Q$. In particular, we are interested in
this function over the interval of $\nu_1$-values that parameterizes
the stability boundary in the $(j_{1r},\gamma)$--plane in the case
where $j_{1r}$ is allowed to vary independently of $j_{1i}$, which is
held fixed with $j_{1i}\Delta t\ne 2n\pi$
in~(\ref{eq:j1rpar}-\ref{eq:g1par}).  Our first lemma treats the case
where $j_{1i}\Delta t\ne n\pi$ for any $n\in Z$, while the
second lemma describes the degenerate case $j_{1i}\Delta t=(2n+1)\pi$
for some $n\in Z$. Each lemma consists of two parts: the first part
merely defines the interval of $\nu_1$--values that parameterizes the
stability boundary in each $(j_{1r},\gamma)$--plane ($j_{1i}$ fixed)
and the second part describes $w(\nu_1,Q)$ on that interval. The first
part of each lemma follows directly from the analysis of Reddy {\it et
  al.}~\cite{RSJ2000}, described in Section~\ref{sec:background}.

\begin{lemma}
\label{tm:lists1}
Consider $\gamma<0$ and let $j_{1r}$
in~(\ref{eq:j1rpar}-\ref{eq:g1par}) vary independently
of $j_{1i}$, which is held fixed with $j_{1i}\Delta t\ne n\pi $
for any $n\in Z$.  (1) The stability boundary in the
$(j_{1r},\gamma)$--plane is associated with an interval
$I_{\nu}=(\nu_a,\nu_b)$ of $\nu_1$-values.  The endpoints, $\nu_a$ and
$\nu_b$, are $j_{1i}\Delta t$ and $2\pi n$, where $n\in Z$ is chosen
so that $|j_{1i}\Delta t-2\pi n|<\pi$.  (2) For $\nu_1\in I_{\nu}$ the
function $w(\nu_1,Q)$ (with $Q$ fixed) 
is positive with $w(\nu_1,Q)\rightarrow 0$ as
$\nu_1$ approaches $\nu_a$ and $\nu_b$; $w(\nu_1,Q)$ has a unique
maximum value $w_{max}(Q)$ at $\nu_1=\nu^*$, where
\begin{equation}
\sin(\nu^*)=j_{1i}(Q)\Delta t-\nu^*\ ,\quad
w_{max}(Q)=1-\cos(\nu^*)\ .
\label{eq:wmaxnu*}
\end{equation}
\end{lemma}

\begin{figure}
  \centerline{\resizebox{3in}{!}{\includegraphics{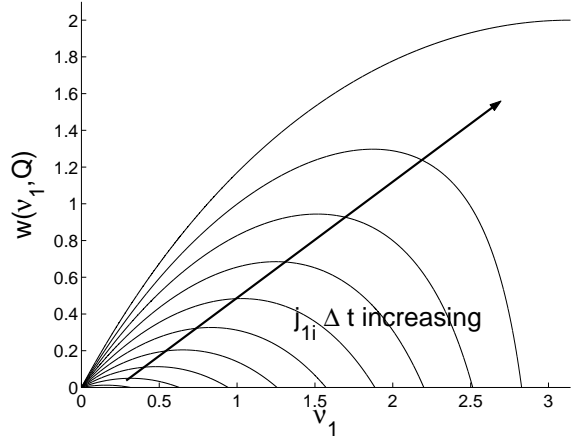}}}
\caption{Plots of $w(\nu_1,Q)$ {\it vs.} $\nu_1$, where $w$ is defined
by~(\ref{eq:QV2}), for various (fixed) values of $Q$ such that
$j_{1i}\Delta t\in(0,\pi]$. 
}
\label{fig:paramchange}
\end{figure}

\begin{proof} {F}rom Reddy~{\it et al.}~\cite{RSJ2000}, we have that 
the stability boundary (for $\gamma<0$) extends between the
  origin of the $(j_{1r},\gamma)$--plane and the asymptote
  $\gamma=j_{1r}$ as $\gamma\to -\infty$. The boundary is associated
  with an open interval $I_{\nu}$; the endpoints of the interval
  correspond to $j_{1i}\Delta t$ (at the origin) and the nearest
  value of $2 \pi n$ (as $\gamma\rightarrow -\infty$). In particular,
  the following two possibilities follow from the
  parameterization~(\ref{eq:j1rpar}-\ref{eq:g1par}): (a) there exists
  an $n\in Z$ such that $j_{1i}\Delta t-2\pi n< \pi$ in which case
  $I_{\nu}=(2\pi n,j_{1i}\Delta t)$, or (b) there exists an $n\in Z$
  such that $2\pi n-j_{1i}\Delta t< \pi$ in which case
  $I_{\nu}=(j_{1i}\Delta t,2\pi n)$. In both cases it is straightforward
  to show that $w(\nu,Q)$, given by~(\ref{eq:QV2}), is positive on the
  interval, approaching 0 at the endpoints $2\pi n$ and $j_{1i}\Delta
  t$.

For fixed $Q$, we find the maximum of $w(\nu_1,Q)$ on $I_{\nu}$
by examining
\begin{equation}
\frac{\partial w(\nu_1,Q)}{\partial \nu_1}
=\frac{(j_{1i}(Q) \Delta t-\nu_1-\sin(\nu_1))} 
{(1+\cos(\nu_1))} \ .
\label{eq:first}
\end{equation}
For $\nu_1 \in I_{\nu}$, the denominator on the right-hand-side
of~(\ref{eq:first}) is positive and the numerator is a decreasing
function of $\nu_1$ which goes through zero at $\nu_1=\nu^*$, where
$\nu^*$ is defined implicitly by~(\ref{eq:wmaxnu*}). The solution $\nu^*$
of $\frac{\partial w}{\partial\nu_1}=0$ is unique and corresponds to a
maximum of $w$ on the interval $I_{\nu}$. In particular, it follows
from~(\ref{eq:first}) that $w(\nu_1,Q)$ is increasing for
$\nu_1\in(\nu_a,\nu^*)$ and decreasing for $\nu_1\in(\nu^*,\nu_b)$.
(See Figure~\ref{fig:paramchange} for sample plots of $w(\nu_1,Q)$ on
$I_\nu$).
\qquad\end{proof}

\begin{lemma}
\label{tm:lists2}
Consider $\gamma<0$ and let $j_{1r}$
in~(\ref{eq:j1rpar}-\ref{eq:g1par}) vary independently of $j_{1i}$,
which is held fixed with $j_{1i}\Delta t=(2n+1) \pi $ for some $n\in
Z$. (1) The stability boundary in the $(j_{1r},\gamma)$--plane
consists of the line segment $\gamma=-j_{1r}$ for
$\gamma\in(0,-\frac{1}{\Delta t}]$ and the parameterized
curve~(\ref{eq:j1rpar}-\ref{eq:g1par}) obtained (equivalently) with
either the interval $I_{\nu}^-\equiv (2n\pi,(2n+1)\pi)$ or
$I_\nu^+\equiv ((2n+1)\pi,2(n+1)\pi)$. (2) For $\nu_1\in I_\nu^\pm$
$w(\nu_1,Q)$ is a positive function, with $w\to 0$ as $\nu_1\to 2m\pi$
($m=n$, $m=(n+1)$ for $I_\nu^-$, $I_\nu^+$, respectively), and $w$
approaches its maximum value $w_{max}=2$ as $\nu_1\to (2n+1)\pi$. The
function $w(\nu_1,Q)$ on $I_\nu^-$ is mapped to $w(\nu_1,Q)$ on
$I_\nu^+$ under the reflection $\nu\to 2(2n+1)\pi-\nu$.  It is an
increasing function on $I_\nu^-$ and hence a decreasing function on
$I_\nu^+$.
\end{lemma}

\begin{proof}
(1) In the degenerate case $j_{1i}\Delta t=(2n+1)\pi$, the line
$\gamma=-j_{1r}$ is a critical curve which forms the stability
boundary for $\gamma\ge -\frac{1}{\Delta t}$; see
Fig.~\ref{fig:degenerate}(a).  This part of the stability boundary
meets another critical curve at $j_{1r}=-\gamma=\frac{1}{\Delta t}$;
this follows from considering~(\ref{eq:j1rpar}-\ref{eq:g1par}) in
the limit  $\nu_1\to (2n+1)\pi$. The latter critical curve extends
to $j_{1r},\gamma\to -\infty$ along $\gamma=j_{1r}$ in the limit
$\nu_1\to 2n\pi,2(n+1)\pi$. That the $\nu_1$--intervals $I_\nu^-$
and $I_\nu^+$ map out the same curve in the $(j_{1r},\gamma)$--plane
follows from the reflection symmetry
of~(\ref{eq:j1rpar}-\ref{eq:g1par}) under $\nu\to 2j_{1i}\Delta
t-\nu$ for $j_{1i}\Delta t=(2n+1)\pi$.  This symmetry is also
necessarily manifest in $w(\nu_1,Q)$.

(2) The proof of the other claims about $w(\nu_1,Q)$ on $I_\nu^\pm$
for $j_{1i} \Delta t=(2n+1)\pi$ are straightforward. We note that the
monotonicity of $w(\nu_1,Q)$ on $I_\nu^\pm$ is determined
by~(\ref{eq:first}).  In particular, for $j_{1i}\Delta t=(2n+1)\pi$,
there is a unique zero of ${\partial w}\over{\partial \nu_1}$ at
$\nu_1=(2n+1)\pi$, which corresponds to a maximum of $w(\nu_1,Q)$ on
the interval $(2n\pi,2(n+1)\pi)$. Thus $w(\nu_1,Q)$ is an increasing
function on $I_\nu^-$, while it is decreasing on the interval $I_\nu^+$.
\qquad\end{proof}

We may now appreciate the relationship between $\nu_1$ and $Q$ along the
stability boundary by considering the following procedure: for each
$Q$ for which $f(Q)> 0$ calculate $j_{1i}(Q)$ and $f(Q)$ and then plot
the left and right hand sides of~(\ref{eq:QV1}) as functions of
$\nu_1$ over the appropriate interval $I_\nu$ to determine which (if
any) $\nu_1$ values correspond to each value of $Q$. The
left--hand--side of~(\ref{eq:QV1}) yields a horizontal line with
positive intercept. If $j_{1i}(Q)\Delta t\neq n\pi$, then by
Lemma~\ref{tm:lists1}, the right--hand--side of~(\ref{eq:QV1}), given
by~(\ref{eq:QV2}), is a positive function with a unique maximum (and
no other critical points) over $I_{\nu}$; see
Fig.~\ref{fig:paramchange} for an example.  Hence, in this case, there
are either 0, 1, or 2 values of $\nu_1$ associated with each $Q$.
In this nondegenerate case we characterize the number of
intersections of the curves associated with the left-- and
right--hand--sides of~(\ref{eq:QV1}) over the interval $I_{\nu}$ 
by determining whether the following distance
function is positive (no intersections), zero (1 intersection) or
negative (two intersections):
\begin{equation}
D(Q)=f(Q)\Delta t-w_{max}(Q)\ ,
\label{eq:distance}
\end{equation}
where $w_{max}(Q)$ is given by~(\ref{eq:wmaxnu*}). We note that $D(Q)$
is continuous. This follows from the continuity of $f(Q)$ and
$j_{1i}(Q)$.  In particular, since $j_{1i}(Q)$ is continuous, the
unique solution $\nu^*$ of~(\ref{eq:wmaxnu*}) is continuous in $Q$,
thereby ensuring the continuity of $w_{max}(Q)$.

For the degenerate cases prescribed by $Q=Q_m$, where $Q_m$ is defined
by $j_{1i}(Q_m)\Delta t= m\pi$, we define
\begin{equation}
w(\nu_1=m\pi,Q_m)\equiv\lim_{\nu_1\to m\pi}w(\nu_1,Q_m)=1-(-1)^{m}\ .
\end{equation}
This corresponds to $w_{max}(Q_m)$, which is $2$ for odd $m$ and $0$
for even $m$.  We now address the case that $m$ is odd, {\it i.e.}
$m=2n+1$ for some $n\in Z$, in which case $w(\nu_1,Q_{2n+1})$ is a
positive function on the interval $(2n\pi,2(n+1)\pi)$, and it is
symmetric about $\nu_1=(2n+1)\pi$.  Thus, for $Q=Q_{2n+1}$ defined by
$j_{1i}(Q_{2n+1})\Delta t=(2n+1)\pi$, it follows from
Lemma~\ref{tm:lists2} that there are either no solutions or one
solution of~(\ref{eq:QV1}) for $\nu_1\in(2n\pi,(2n+1)\pi]$ or,
equivalently, for $\nu_1\in[(2n+1)\pi,(2n+2)\pi)$. We recall that in
this degenerate case there is always an additional point on the
stability boundary for $\gamma=-f(Q_{2n+1})/2$ provided
$0<f(Q_{2n+1})\Delta t<2$; this point lies at the intersection of the
lines $j_{1r}=-\gamma$ and $j_{1r}=\gamma+f(Q_{2n+1})$ (see part (1)
of Lemma~\ref{tm:lists2}).  Moreover, the condition
$0<f(Q_{2n+1})\Delta t<2$ is equivalent to $f(Q_{2n+1})\Delta
t<w_{max}(Q_{2n+1})$ since $w_{max}=2$ in this case.  Thus the
interpretation of the sign of $D(Q)$, defined by~(\ref{eq:distance}),
is the same in this degenerate case as it is in the nondegenerate
case.  Specifically, if $D(Q_{2n+1})<0$ then there are two points on
the stability boundary associated with this value of $Q_{2n+1}$; if
$D(Q_{2n+1})=0$, then there is a single point on the stability
boundary (given by $(j_{1r},\gamma)=(1/\Delta t,-1/\Delta t)$); and if
$D(Q_{2n+1})>0$, then there are no points on the stability boundary
associated with this value of $Q_{2n+1}$.

The next lemma addresses the continuity of particular critical curves
that are parameterized by $Q$. It is specifically concerned with the
case that $j_{1i}(Q) \Delta t =(2n+1)\pi$ ($n\in Z$) at some point
along the curve. At such points the relevant interval $I_\nu$ of
$\nu_1$--values may ``jump'' ({\it e.g.} from $I_\nu^{-}$ to
$I_\nu^{+}$ in the case that $Q=Q_{2n+1}$).  However, this does not lead to
a corresponding discontinuity in the critical curve parameterized by
$Q$, as we now show.

\begin{lemma}
Consider a critical curve in the $(j_{1r},\gamma)$--plane,
parameterized by $Q$, which extends from the origin to $\gamma\to -\infty$
along the asymptote $\gamma=j_{1r}$. 
The critical curve is continuous at any points along it that are
parameterized by $Q=Q_{2n+1}$, where $j_{1i}(Q_{2n+1})\Delta t=(2n+1)\pi$ for
some integer $n$, and $f(Q_{2n+1})\Delta t\in (0,2)$.
\label{tm:continuous}
\end{lemma} 
  
\begin{proof}
  Let $Q_{2n+1}$ be in the set of $Q$ values that parameterizes the
  critical curve in the $(j_{1r},\gamma)$--plane, where $Q_{2n+1}$ is
  defined by the condition $j_{1i}(Q_{2n+1})\Delta t=(2n+1)\pi$ for some
  integer $n$. Provided $f(Q_{2n+1}) \Delta t\in (0,2)$, 
then there are two points in the $(j_{1r},\gamma)$-plane, denoted
  $P_1$ and $P_2$, associated with $Q_{2n+1}$. 
(See part (1) of Lemma~\ref{tm:lists2}.) 
Specifically, let $P_1$  be the
point that is associated with the unique solution
  of~(\ref{eq:QV1}-\ref{eq:QV2}) on $I_\nu^-=(2n\pi,(2n+1)\pi)$ (or,
  equivalently, on $I_\nu^+=((2n+1)\pi,(2n+2)\pi)$, and  let $P_2$ be the point
that lies at the intersection of the lines $j_{1r}=-\gamma$ and
$j_{1r}=f(Q_{2n+1})+\gamma$. 
Since $j_{1i}(Q)\Delta t$ varies continuously with $Q$, then for values of
$Q$ in a neighborhood of $Q_{2n+1}$ we expect the interval $I_\nu$
of $\nu_1$--values associated with the 
mapping~(\ref{eq:QV1}-\ref{eq:QV2}) to be as described in 
Lemma~\ref{tm:lists1}.
For example, if $\frac{dj_{1i}}{dQ}(Q_{2n+1})>0$, 
then $j_{1i}(Q) \Delta t$ increases through $(2n+1)\pi$ as $Q$  
increases through $Q_{2n+1}$ and the relevant interval of $\nu_1$--values
switches from $(2n\pi,j_{1i}(Q)\Delta t)$ to $(j_{1i}(Q)\Delta
t,2(n+1)\pi)$ as $Q$ increases through $Q_{2n+1}$. 
In any case, for $Q$ sufficiently close to $Q_{2n+1}$ so
  that $f(Q)\Delta t$ remains smaller than $w_{max}(Q)$, there are two
  solutions $\nu_1$ of~(\ref{eq:QV1}-\ref{eq:QV2}) on the appropriate
  $I_\nu$--interval.  These values of $\nu_1$ determine a pair of points on
  the critical curve in the $(j_{1r},\gamma)$-plane via the parametric
  equations~(\ref{eq:j1rpar}-\ref{eq:g1par}). We now show that these points
converge to $P_1$ and $P_2$ as $Q\to Q_{2n+1}$. Specifically, 
  one of the pair of solutions is associated with a $\nu_1$--value in the
  the interval $(2n\pi,\nu^*)\subset I_\nu^-$ if $j_{1i}(Q) \Delta t
  <(2n+1)\pi$, and with the interval $(\nu^*,(2n+2)\pi)\subset I_\nu^+$
  if $j_{1i}(Q) \Delta t>(2n+1)\pi$, where $\nu^*$ determines
  $w_{max}$ via~(\ref{eq:wmaxnu*}). (This is the $\nu_1$ solution that is
furthest from $(2n+1)\pi$.) This $\nu_1$-value determines
a  point in the $(j_{1r},\gamma)$--plane that converges to $P_1$ as
  $Q\to Q_{2n+1}$ since $w(\nu_1,Q)$ changes continuously on that
  subinterval as $Q\to Q_{2n+1}$. 
To see that the
second of the pair of $\nu_1$ solutions describes a point that converges to
  $P_2$ as $Q\to Q_{2n+1}$, we first note that the $\nu_1$ value must lie
  between $\nu^*$ and $(2n+1)\pi$ and that $\nu^*\to (2n+1)\pi$ as
  $Q\to Q_{2n+1}$.  Hence this solution $\nu_1\to (2n+1)\pi$ as 
$Q\to Q_{2n+1}$,
  and the associated point in the $(j_{1r},\gamma)$-plane
must approach the line $j_{1r}=-\gamma$ as $Q\to Q_{2n+1}$
  since $j_{1r}=\gamma \cos(\nu_1)$. Moreover, the value of
  $j_{1r}\equiv f(Q)+\gamma$ converges to $f(Q_{2n+1})+\gamma$ as $Q\to
  Q_{2n+1}$; it then follows that this point converges to $P_2$. Thus we
  have shown that this critical curve in the $(j_{1r},\gamma)$-plane,
  parameterized by $Q$, is continuous at any points $Q=Q_{2n+1}$ associated
with the parameterization.  
\qquad
\end{proof}

The following lemma provides a sufficient condition for a
nondegenerate critical curve to lie inside of the existence boundary
for fixed values of $b_1$, $b_3$, $K$, and $\rho$, {\it i.e.}  as in
Fig.~\ref{fig:outbound}(b). Thus, it determines a sufficient condition
for stabilization via time--delay feedback to fail.

\begin{lemma}
  Assume that there is a $Q$--interval $I_Q=[\widehat Q_1,\widehat
  Q_2]$ on which $j_{1i}(Q) \Delta t \ne 2n\pi$ for any $n\in Z$ and
  on which $f(Q)\ge 0$ with $f(Q)=0$ only at the endpoints $\widehat Q_1$ and
  $\widehat Q_2$.  If there exists a value $Q\in I_Q$ for which $D(Q)=0$, where
  $D(Q)$ is given by (\ref{eq:distance}), then the traveling wave
  cannot be (linearly) stabilized for any value of the time delay
  feedback parameter $\gamma$.
\label{tm:unstab}
\end{lemma} 

\begin{proof}
Assume that the conditions of the lemma are met and let $Q^*$ be the
largest value of $Q\in I_Q$ for which $D(Q)=0$. That the traveling
wave cannot be stabilized for $\gamma\ge 0$ follows from part (b) of
Lemma~\ref{lemma-fpos}. For $\gamma <0$ we show below that there is
a continuous critical curve, parameterized by $Q\in [Q^*,\widehat Q_2]$, that
extends from $\gamma=j_{1r}=0$ to $\gamma\to -\infty$ along the
asymptote $j_{1r}=\gamma$.  

To see that there is a continuous critical curve parameterized by
$Q\in [Q^*,\widehat Q_2]$, we must consider the
mapping~(\ref{eq:QV1}-\ref{eq:QV2}) between $\nu_1$ and $Q$ over this
range of $Q$--values.  Since $D(Q)$ is continuous, with $D(Q)=0$ only
at the endpoint $Q=Q^*$, then the sign of $D(Q)$ for
$Q\in(Q^*,\widehat Q_2]$ is the same as the sign of $D(\widehat Q_2)$.
Since $w_{max}(Q)$, defined by~(\ref{eq:wmaxnu*}), is positive
whenever $j_{1i}(Q)\Delta t\ne 2n\pi$, and since $f(\widehat Q_2)=0$,
we know that $D(\widehat Q_2)<0$.  While for $Q=Q^*$ there is a unique
value $\nu_1=\nu^*$ associated with the
mapping~(\ref{eq:QV1}-\ref{eq:QV2}), which is given
by~(\ref{eq:wmaxnu*}), there are two corresponding values of $\nu_1$
for all other values of $Q$ in the interval since $D(Q)<0$ ({\it i.e.}
for $Q\in(Q^*,\widehat Q_2)$). (If ever $j_{1i}(Q)\Delta t=(2n+1)\pi$,
then one of these $\nu_1$ values is $(2n+1)\pi$.)  All of the values
of $\nu_1$ lie in a $\nu$--interval $J_{\nu}=(2n\pi ,2(n+1)\pi)$ for
some $n$ since $j_{1i}(Q) \Delta t \ne 2m\pi$ for any $m\in Z$ and any
$Q\in [\widehat Q_1, \widehat Q_2]$.  We may now define two functions
$\nu_1^a(Q)$ and $\nu_1^b(Q)$ using the mapping on $[Q^*,\widehat
Q_2]$.  We choose $\nu_1^a(Q)$ to be the solution
of~(\ref{eq:QV1}-\ref{eq:QV2}) that is furthest from the midpoint of
$J_\nu$ ({\it i.e.} furthest from $\nu_1=(2n+1)\pi$) and $\nu_1^b(Q)$ to
be the solution that is closest to $\nu_1=(2n+1)\pi$. We note that the
function $\nu_1^a(Q)$ is discontinuous at any $Q$ where
$j_{1i}(Q)\Delta t$ goes through $(2n+1)\pi$ due to the jumps in the
interval $I_\nu$ from $I_\nu^+$ to $I_\nu^-$.  (In particular, the
discontinuity corresponds to $\nu_1^a$ being reflected to
$2(2n+1)\pi-\nu_1^a$; see Lemma~\ref{tm:lists2}.)  In contrast, there are
no discontinuities in $\nu_1^b(Q)$; it takes on the value $(2n+1)\pi$
at any $Q$ value where $j_{1i}(Q) \Delta t=(2n+1)\pi$.  See
Fig.~\ref{fig:blah} for an example of functions $\nu_1^a(Q)$ and
$\nu_1^b(Q)$ in the discontinuous case. Note that the two functions
approach the unique value $\nu_1=\nu^*$ as $Q\to Q^*$.

\begin{figure}
\centerline{\resizebox{4in}{!}{\includegraphics{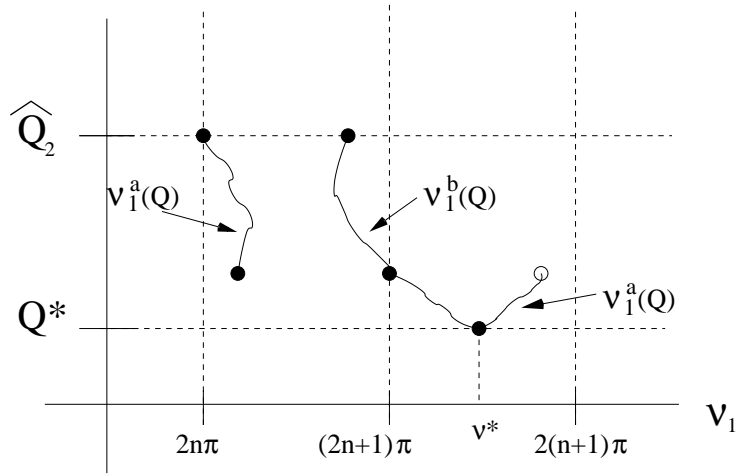}}}
\caption{Schematic of the mappings between $\nu_1$ and $Q$ that
parameterize a critical curve in the $(j_{1r},\gamma)$-parameter
plane. The parameterization by $Q\in[Q^*,\widehat Q_2]$ is determined by the
two functions $\nu_1^a(Q)$ and $\nu_1^b(Q)$, defined in the proof of
Lemma~\ref{tm:unstab}, where $\nu_1\in[2n\pi,2(n+1)\pi]$. The schematic
is typical of the degenerate case where $j_{1i}(Q)\Delta t=(2n+1)\pi$
for some $Q\in(Q^*,\widehat Q_2)$, in which case $\nu_1^a$ is discontinuous
as shown.}
\label{fig:blah}
\end{figure}

A $Q$-parameterized critical curve is obtained by inserting the pairs
$(Q,\nu_1^a(Q))$ and $(Q,\nu_1^b(Q))$ into the parametric equations
for $j_{1r}$ and $\gamma$ given by~(\ref{eq:j1rpar})-(\ref{eq:g1par}).
The continuity of the critical curve associated with $\nu_1^b(Q)$
follows directly from the continuity of the mapping in this case.  The
function $\nu_1^a(Q)$ may be a discontinuous function at isolated
points $Q$ where $j_{1i}(Q)\Delta t=(2n+1)\pi$, as described above.
However, by Lemma~\ref{tm:continuous}, these points do not lead to
corresponding discontinuities in the critical curve. Thus the portion
of the critical curve parameterized by $(Q,\nu_1^a(Q))$ is also
continuous. The two functions together parameterize a continuous
critical curve from the origin in the $(j_{1r},\gamma)$--plane to
$\gamma\to -\infty$ along $\gamma=j_{1r}$. This follows since the two
curves meet as $Q\to Q^*$, and since
$(\widehat Q_2,\nu_1^b(\widehat Q_2))=(\widehat Q_2,j_{1i}(\widehat Q_2) \Delta t )$ and
$(\widehat Q_2,\nu_1^a(\widehat Q_2))=(\widehat Q_2,2\pi m)$, where $m=n$ or $m=(n+1)$. Thus the
curve parameterized by $(Q,\nu_1^b(Q))$ approaches the origin of the
$(j_{1r},\gamma)$-plane as $Q\to \widehat Q_2$ and the curve parameterized by
$(Q,\nu_1^a(Q))$ approaches, as $Q\to \widehat Q_2$, the asymptote
$\gamma=j_{1r}$ with $\gamma\to -\infty$.\qquad \end{proof}

We now use the results of the previous lemmas to prove our main
stability result stated in Theorem \ref{tm:characteristic}.

\begin{proof} {\it of Theorem~\ref{tm:characteristic}.}  
The necessary condition for  linear stability of
  the traveling wave follows from Lemma~\ref{tm:unstab} in the case that
$j_{1i}(Q)\Delta t\ne 2\pi n$ for any $n\in Z$ and for any $Q$ for which
$f(Q)\ge 0$. 
 Note, in particular, that the
  equations~(\ref{eq:crit1}-\ref{eq:crit2}) in
  Theorem~\ref{tm:characteristic} are equivalent to the condition
  $D(Q)=0$ in Lemma~\ref{tm:unstab}, through the definition of
  $w_{max}(Q)$ given by~(\ref{eq:wmaxnu*}).
  
  If $j_{1i}(Q)\Delta t =2\pi n$ for some $Q=Q_{2n}$ where $f(Q_{2n})\ge 0$,
  then Lemma~\ref{tm:degeneracies} states that the traveling wave
  cannot be made linearly stable (with the prescribed value of $\rho$)
  for any value of $\gamma$ associated with the time-delay feedback.
  We now need to show that in this
  case~(\ref{eq:crit1}-\ref{eq:crit2}) has a solution. If $f(Q_{2n})=0$,
  then $(Q,\nu_1)=(Q_{2n},2\pi n)$
  solves~(\ref{eq:crit1}-\ref{eq:crit2}).  If $f(Q_{2n})>0$, then we can
  show there must be a solution to $D(Q)=0$ on the interval
  $(Q_{2n},\widehat Q_2)$, where $\widehat Q_2$ is the zero of $f(Q)$ that is 
closest to
  $Q_{2n}$, {\it i.e.}  $f(\widehat Q_2)=0$ and $f(Q)>0$ for all $Q\in(Q_{2n},\widehat Q_2)$.
  (Recall that the equation $D(Q)=0$, where $D(Q)$ is defined
  by~(\ref{eq:distance}), is equivalent
  to~(\ref{eq:crit1}-\ref{eq:crit2}).)  To see that there is a
  $Q\in(Q_{2n},\widehat Q_2)$ for which $D(Q)=0$, note that $D(Q_{2n})>0$ since
  $w_{max}(Q_{2n})=0$ and $D(\widehat Q_2)<0$ since $f(\widehat Q_2)=0$. The result follows
  from the continuity of $D(Q)$.
  
  To complete the proof we show that the traveling wave may be
  stabilized at $\gamma=-\frac{1}{\Delta t}$ if there is no solution
  of~(\ref{eq:crit1}-\ref{eq:crit2}). This follows since the
  equations~(\ref{eq:crit1}-\ref{eq:crit2}) describe points on a
  critical curve within the accessible region for
  $\gamma=-\frac{1}{\Delta t}$.
Thus there is no critical
  curve that passes through $\gamma=-\frac{1}{\Delta t}$ if there is
  no solution to~(\ref{eq:crit1}-\ref{eq:crit2}).
In this case, the stability boundary must cross the existence
  line as in Fig.~\ref{fig:outbound}(a), and the traveling wave can be
  stabilized for some range of $\gamma$ values, including $\gamma =
  -\frac{1}{\Delta t}$.\qquad\end{proof}

The significance of Theorem~\ref{tm:characteristic} is that it
determines an ``optimum'' value of $\gamma$ to use in stabilizing the
traveling wave; this is the value $\gamma=-\frac{1}{\Delta t}$. To be
more precise, any set of $\gamma$ values that stabilize the traveling
wave must include $\gamma=-\frac{1}{\Delta t}$.  We note that this
$\gamma$-value coincides with where the stability boundary is furthest
from the asymptote $j_{1r}=\gamma$ for each {\it fixed} value of
$j_{1i}\Delta t\ne 2n\pi$ (see part (1) of
Lemma~\ref{tm:othercurves}). The power of
Theorem~\ref{tm:characteristic} is that it replaces the ``brute force''
approach of determining whether there are any growing solutions of the
linear delay equation~(\ref{eq:deleq1}) for each and every value of
$Q$ for which $f(Q)>0$ with the simpler problem of determining whether
there is a solution to an algebraic equation for $Q$ over the same
interval for which $f(Q)>0$.

\section{Numerical Linear Stability Results}
\label{sec:results}

\begin{figure}
\centerline{$\qquad$ (a): $(b_1,b_3)=(1.0,0.5)$  \hspace{.3 in} 
(b): $(b_1,b_3)=(1.5,1.0)$ \hspace{.3 in} (c): $(b_1,b_3)=(2.0,1.0)$  }
\smallskip
\centerline{\resizebox{2in}{!}{\includegraphics{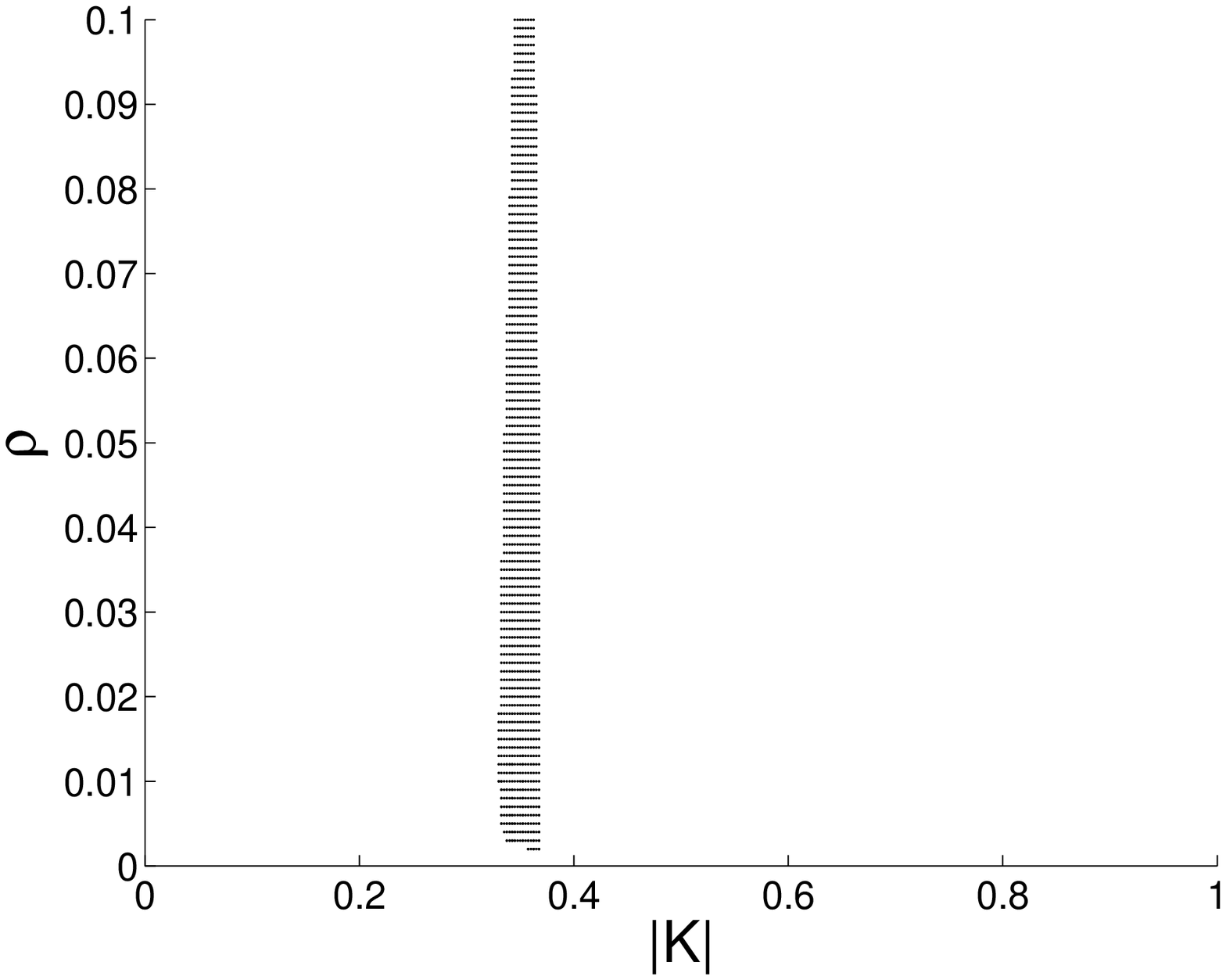}}
\resizebox{2in}{!}{\includegraphics{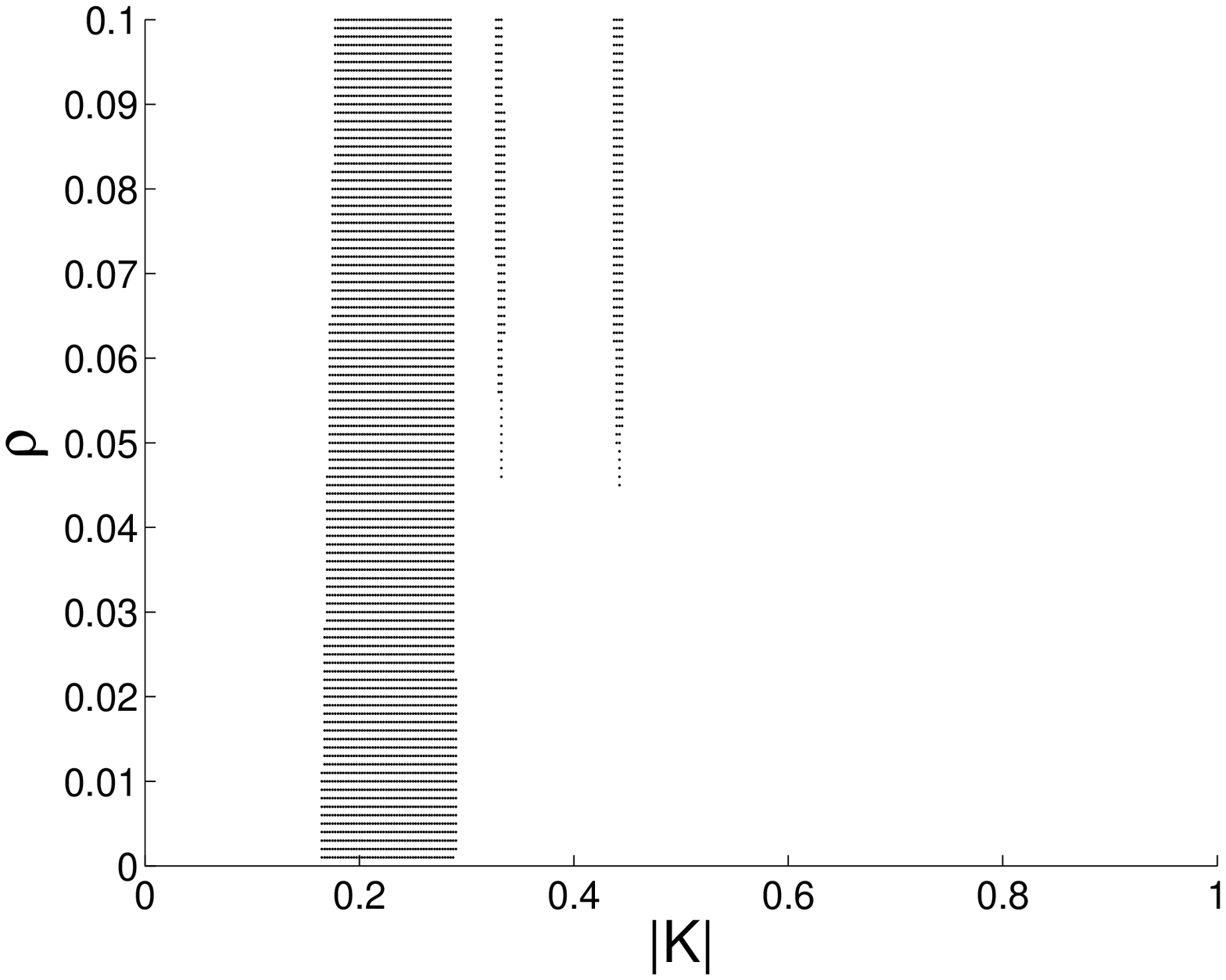}}\resizebox{2in}{!}{\includegraphics{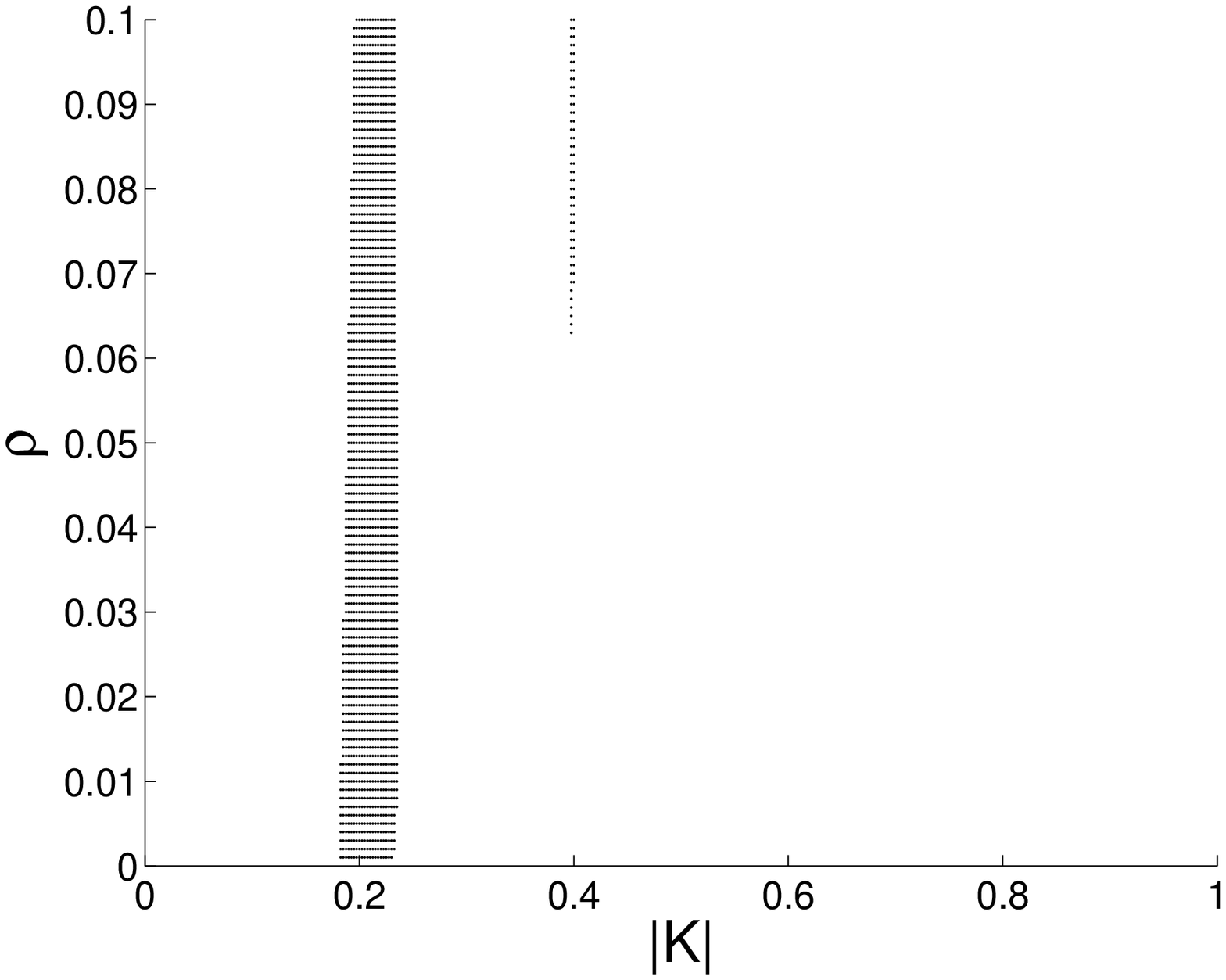}}}
\medskip
\centerline{$\qquad$ (d): $(b_1,b_3)=(2.5,1.0)$  \hspace{.3in} (e): $(b_1,b_3)=(2.0,1.5)$ 
\hspace{.3in} (f): $(b_1,b_3)=(2.5,1.5)$ }
\smallskip
\centerline{\resizebox{2in}{!}{\includegraphics{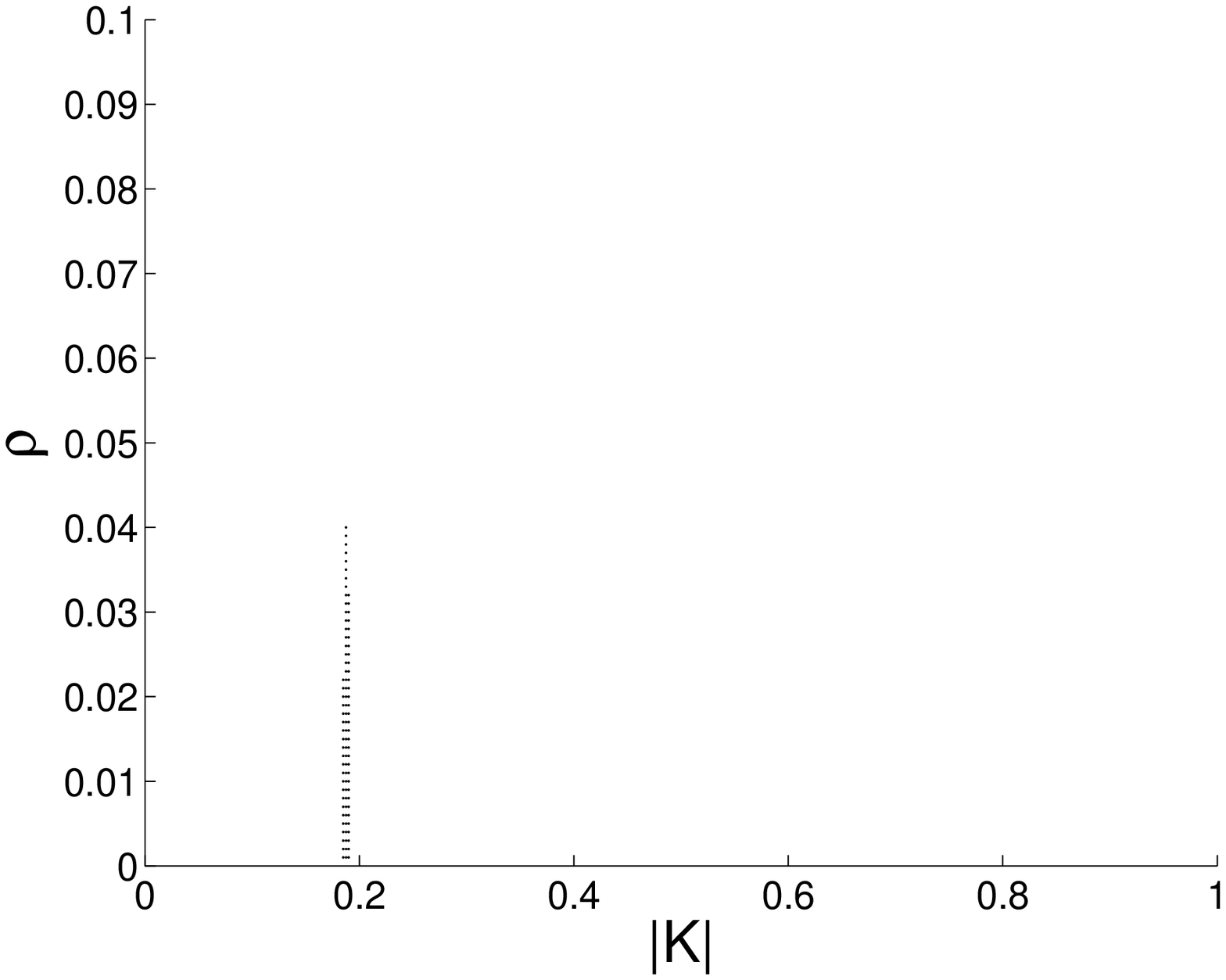}}
\resizebox{2in}{!}{\includegraphics{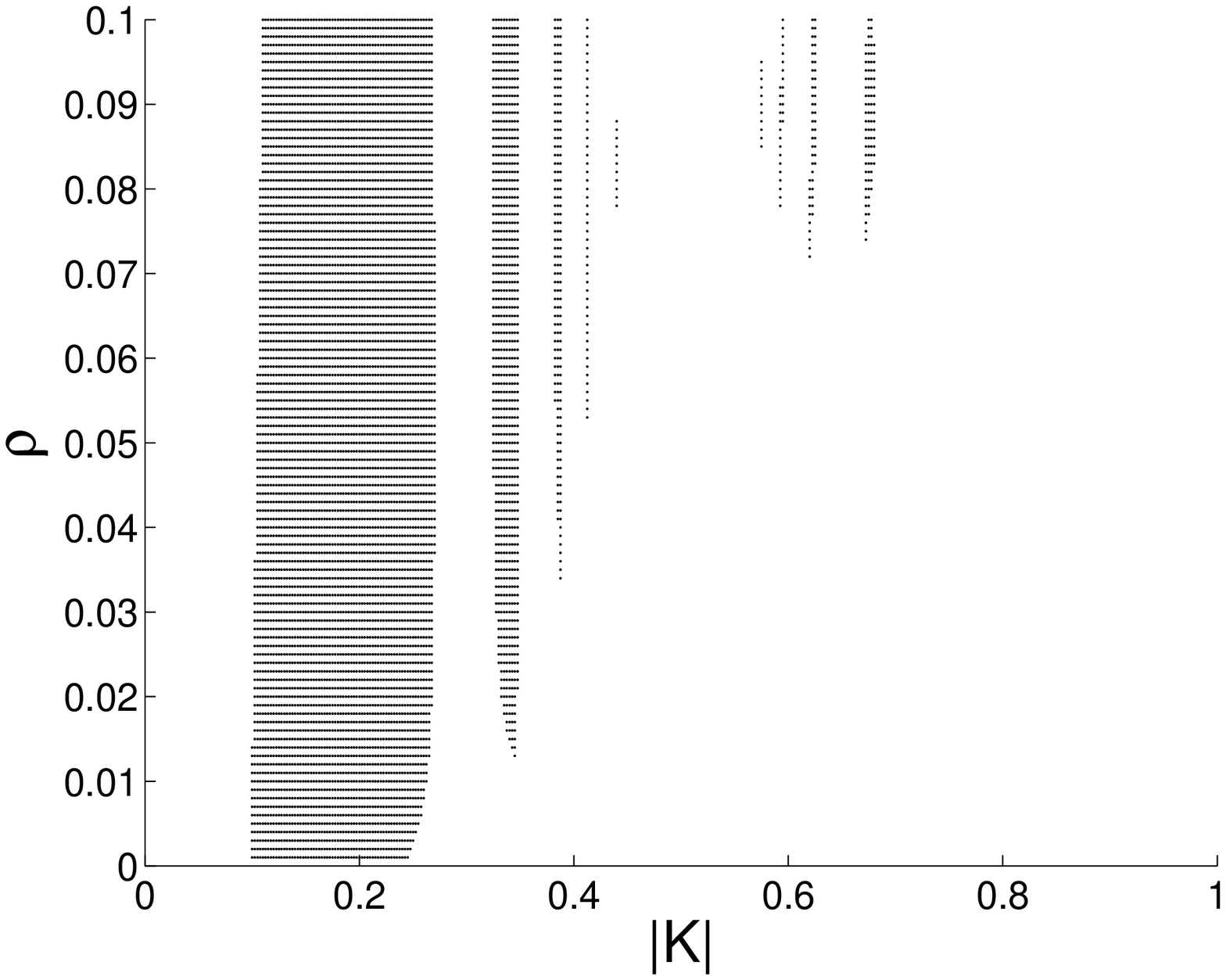}}
\resizebox{2in}{!}{\includegraphics{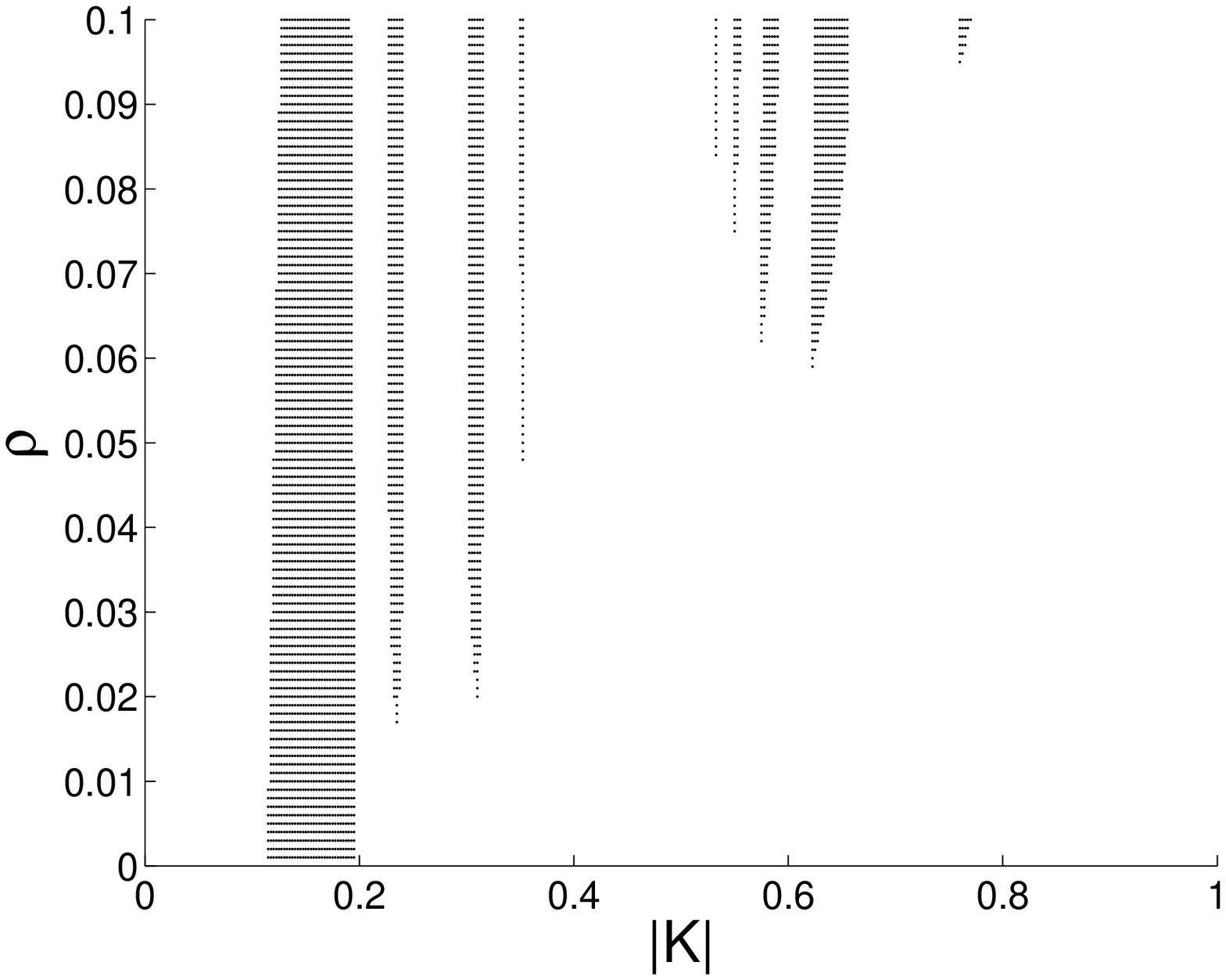}}}
\medskip
\centerline{(g): $(b_1,b_3)=(2.5,2.0)$\hspace{2.3in} (h)$\qquad$ }
\smallskip
\centerline{\resizebox{3in}{!}{\includegraphics{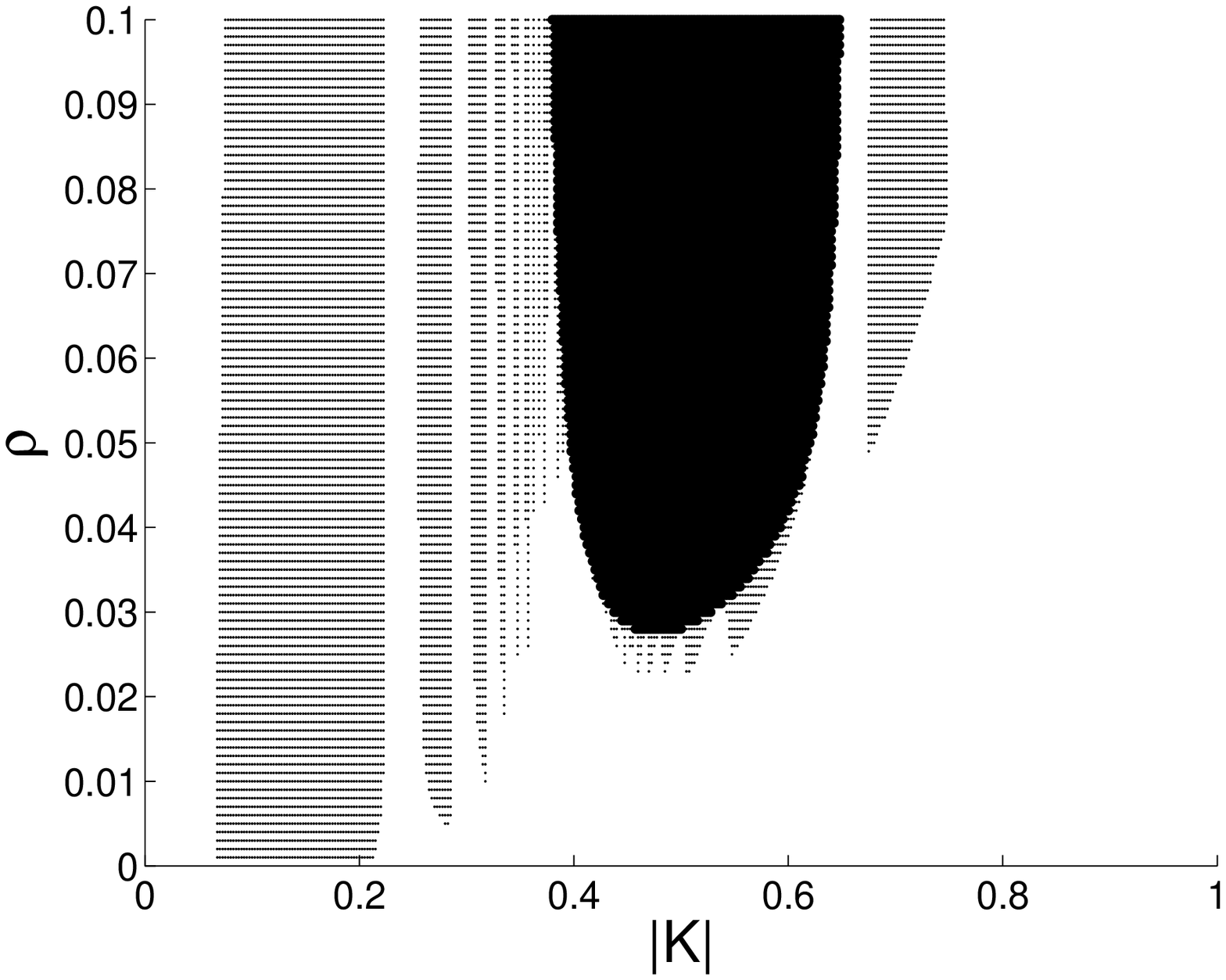}}
\hspace{.3in}
\resizebox{3 in}{!}{\includegraphics{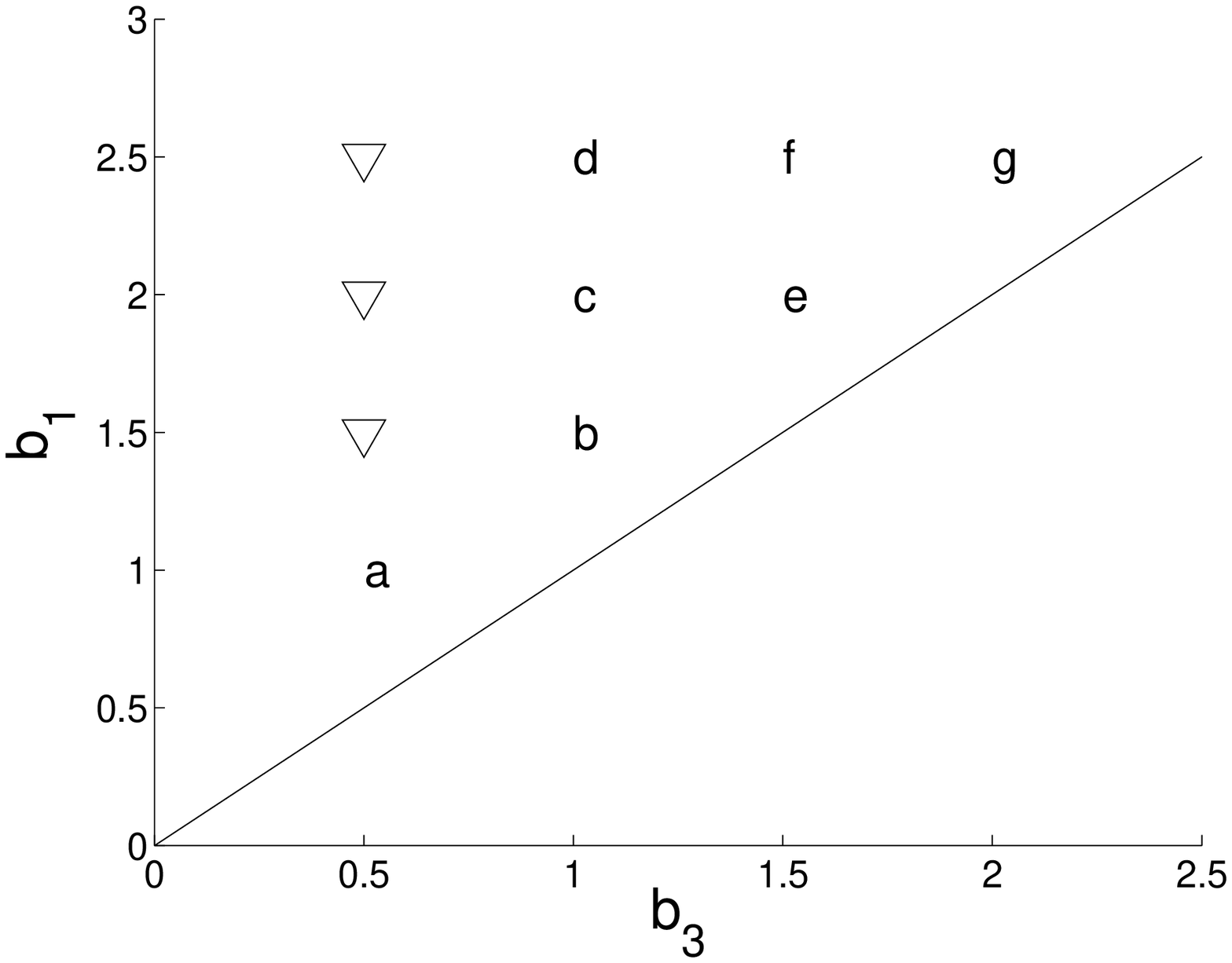}}}
\caption{(a)-(g) Examples of stability diagrams in the
$(|K|,\rho)$-parameter plane, created using the stability criterion of
Theorem~\ref{tm:characteristic}, for the values of $(b_1,b_3)$
indicated above each plot.
The lighter shading indicates regions that can be stabilized by a
combination of spatial and temporal feedback (with $\gamma=-1/\Delta
t$).  In the darker region in (g), the traveling wave can be
stabilized by spatial feedback alone. (h) Summary of our stability
results in the ($b_3,b_1)$--parameter plane.  The letters refer to the
stability diagrams shown in plots (a)-(g), which exhibit stable
regions in the $(K,\rho)$-parameter plane.  At points marked by
triangles, either no stable traveling waves are found, or only very
narrow stable regions near the unphysical limit in which $\Delta t
\rightarrow \infty$.}
\label{fig:stablarge1}
\end{figure}

Figure~\ref{fig:stablarge1} presents numerical linear stability results in the
$(|K|,\rho)$-parameter plane for various values of $(b_1,b_3)$ in the
Benjamin-Feir unstable regime. These diagrams were created using the
stability criterion of Theorem~\ref{tm:characteristic}.  Specifically,
the criterion was applied at points on a grid in the $(K,\rho)$-plane
with spacing of $0.0025$ in the $K$-direction and $0.001$ in the
$\rho$-direction.  In order to determine whether there is a solution
to~(\ref{eq:crit1}-\ref{eq:crit2}) of Theorem~\ref{tm:characteristic},
we first eliminated the variable $\nu$
from~(\ref{eq:crit1}-\ref{eq:crit2}) yielding a single equation for
$Q$. Care was taken with the trigonometric terms in $\nu$, which
needed to be inverted so that the $\nu$ values were in the interval
$I_\nu$ that contained $j_{1i}(Q)\Delta t$ for each value of $Q$. (See
Lemmas~\ref{tm:lists1}-\ref{tm:lists2} for a description of $I_\nu$.)
Newton's method was then used to determine whether there was a
solution for any $Q$ satisfying $ 0 \le f(Q) \Delta t \le 2 $.
Specifically, for each traveling wave tested, the program first
checked all values of $Q$ between .002 and 20 at an increment of .002,
determining the ranges of $Q$ values for which $f(Q)>0$. (Although we
did not prove in all cases that there cannot be an instability for
$Q>20$, this represents an extremely conservative estimate of the
largest $Q$ value for an instability for the $(b_1,b_3,K)$ values
considered; instabilities were generally bounded well below $Q=20$
for the parameters used in the plots.) Next, on each interval for
which $f(Q)\Delta t\in [0,2]$, up to twenty-four $Q$
values were chosen as initial conditions for Newton's method, which
was used to determine whether there was a solution to the stability
criterion equations~(\ref{eq:crit1}-\ref{eq:crit2}) or not.

Note that the black region in Figure~\ref{fig:stablarge1}(g) indicates
parameter values for which $f(Q)<0$ for all $Q$ tested ({\it i.e.} for
$Q\in(0,20)$ at intervals of $.002$). Thus, by Lemma~\ref{lemma-fpos},
the traveling wave can be stabilized by spatially--translated feedback
alone in this region. The lighter shaded regions of
Figure~\ref{fig:stablarge1} indicate parameter values for which the
traveling wave could be stabilized by a combination of spatial and
temporal feedback, at least for the ``optimal choice'' of temporal
feedback parameter $\gamma=-\frac{1}{\Delta t}$.  The feedback failed
to linearly stabilize the traveling wave at all other points in the
parameter plane.

Figure~\ref{fig:stablarge1}(h) summarizes our linear stability results
in the $(b_1,b_3)$-parameter plane; it indicates both parameters for
which we found stable regions in the $(K,\rho)$-plane ({\it cf.}
Figure~\ref{fig:stablarge1}(a)-(g)), and parameter values, marked by
triangles, where either no stable traveling waves were found or only
very narrow stability regions near the unphysical limit in which
$\Delta t \rightarrow \infty$.  From Figure~\ref{fig:stablarge1} it is
evident that the stable regions in the $(K,\rho)$--parameter plane are more
substantial close to the Benjamin-Feir line. Moreover, from a
comparison of Figure~\ref{fig:stablarge1}(h) with
Figure~\ref{fig:phase}, we see that the feedback technique seems to be
more effective in stabilizing traveling waves in the phase turbulent
regime, although it also works for some parameters in the amplitude
turbulent regime.

\begin{figure}
\centerline{\resizebox{2.5 in}{!}{\includegraphics{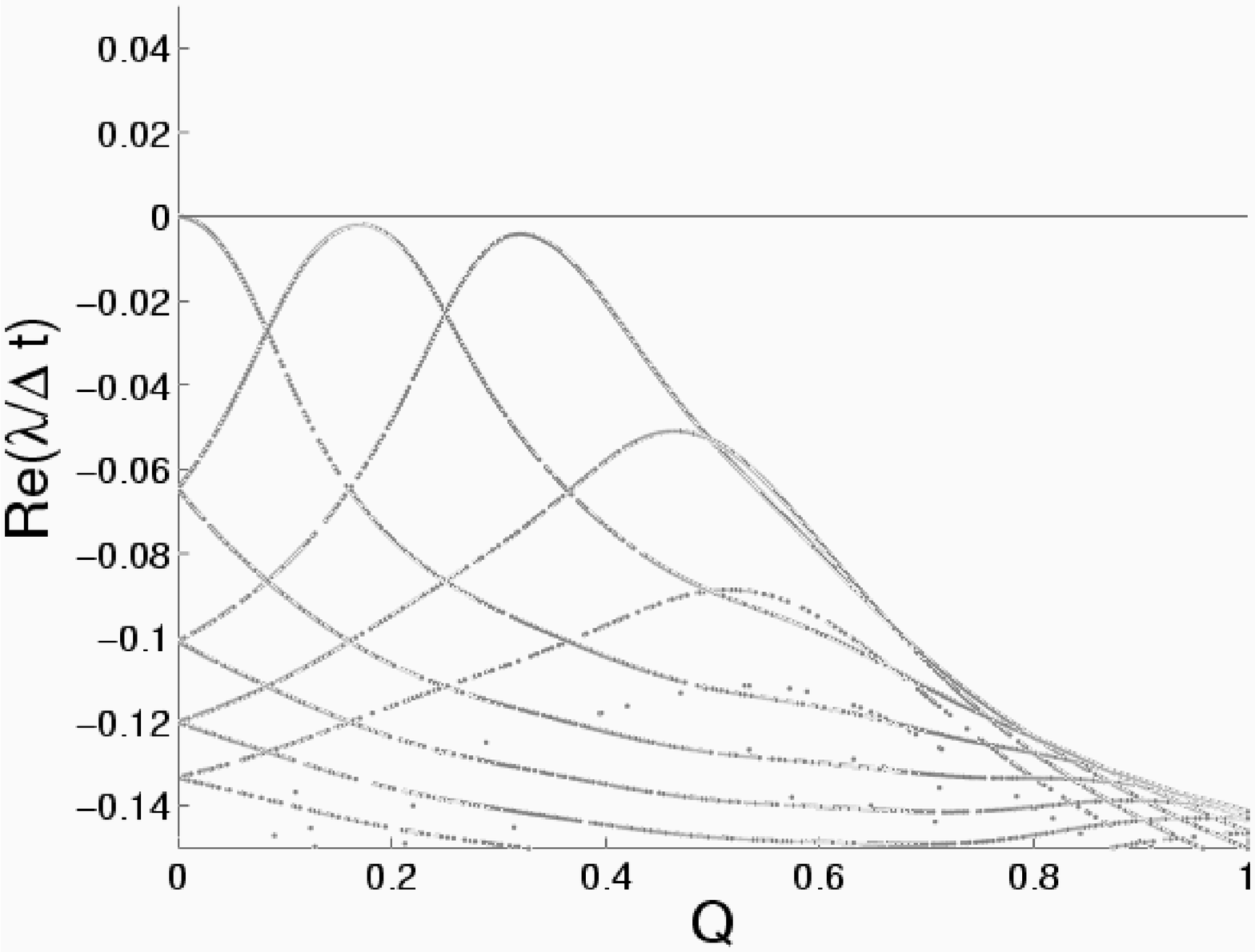}} 
\resizebox{2.5 in}{!}{\includegraphics{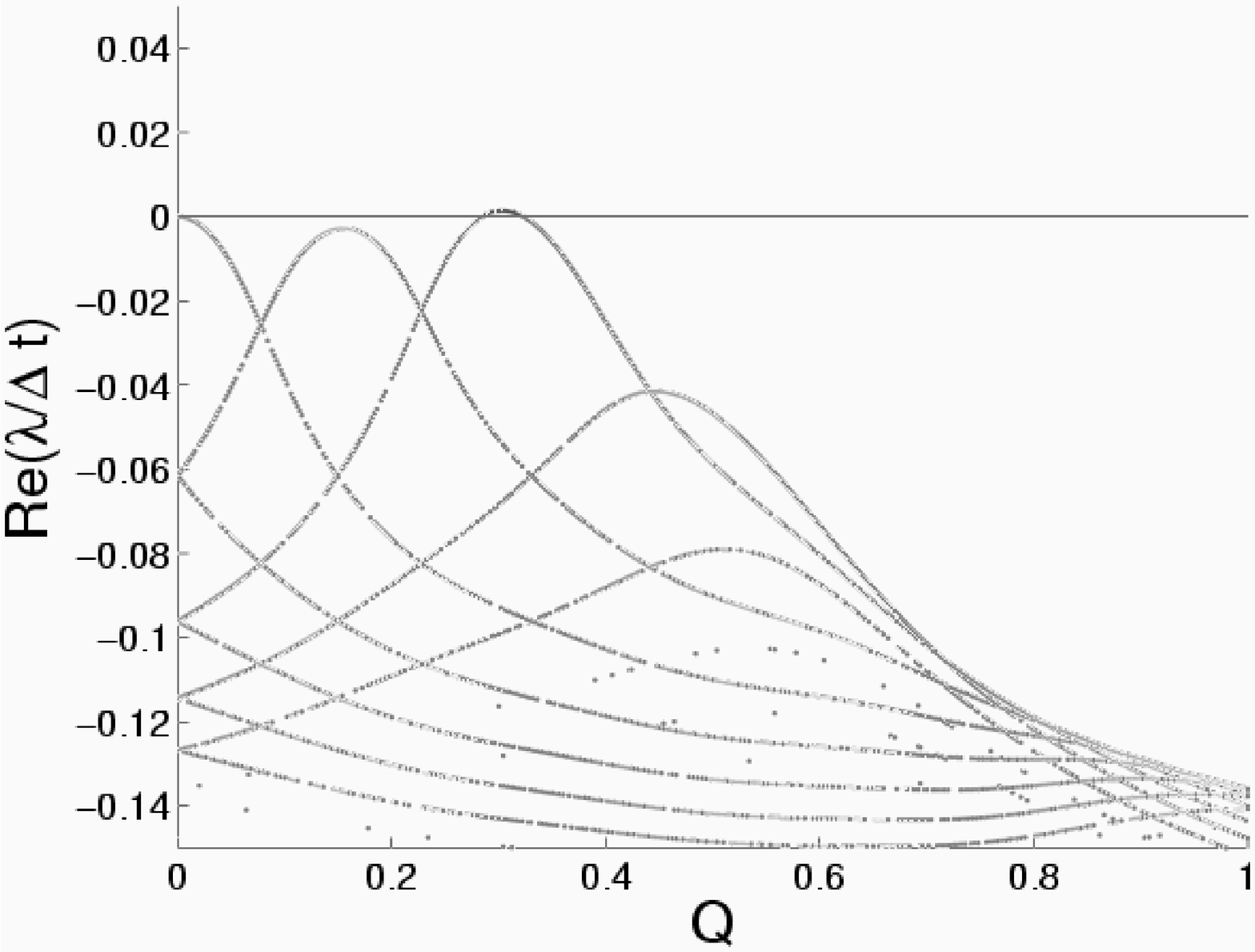}}}
\centerline{$\quad$ (a) \hspace{2.2in} (b)}
\caption{Plots of the growth rates associated with solutions of the
delay equation ~(\ref{eq:deleq1}) (for $j=1$) as a function of $Q$ for
$(b_1,b_3)=(2.5,2.0)$, $\rho=.007$, and (a) $K=.2800$, (b) $K=.2875$.}
\label{fig:growthrates}
\end{figure}

We have performed numerous ``brute force'' checks of the stability
results summarized by Figure~\ref{fig:stablarge1} in order to have
confidence in our numerical implementation of the stability criterion
of Theorem~\ref{tm:characteristic}. An example of such a check is
given in Figure~\ref{fig:growthrates}, which shows the results of our
direct calculation of the growth rates of solutions of the linear
delay equation~(\ref{eq:deleq1}) (for $j=1$) as a function of $Q$ for
two different points in the parameter plane associated with
Figure~\ref{fig:stablarge1}(g).  Figure~\ref{fig:growthrates}(a)
corresponds to a point just inside of the stability boundary, while
Figure~\ref{fig:growthrates}(b) corresponds to a nearby point that is
outside of the stability boundary.

We now examine in greater detail some of the features evident in
Figure~\ref{fig:stablarge1}.  We first note that in each of
Figures~\ref{fig:stablarge1}(b-g) the leftmost stability tongue
extends to $\rho=0$ indicating that the associated $K$--values can be
stabilized with temporal feedback alone.  We find that the $K$--values
associated with the left--most stability tongue in
Figures~\ref{fig:stablarge1}(a)-(g) include a value $K$ for which
spatially--resonant perturbation wavenumbers $\widetilde Q_2= 2K$
have associated frequencies $j_{1i}(\widetilde Q_2)$ that satisfy
$j_{1i}(\widetilde Q_2)\Delta t=-\pi$. The significance of this latter
relation is that it corresponds to the degenerate situation for which
the stability boundary in the $(j_{1r},\gamma)$--plane extends all the
way to the line $\gamma=-j_{1r}$ for $\gamma=-1/\Delta t$; see
Fig.~\ref{fig:degenerate}(a) with $\alpha=j_{1r}$, $G=\gamma$.  In
particular, while perturbations with the resonant wavenumber
$Q=\widetilde Q_2$ are not affected by the spatial feedback, their
growth rates are, loosely speaking and in the sense just described,
maximally reduced by the temporal feedback. These observations suggest
a possible explanation for why this first stability tongue can extend
all the way to $\rho=0$, as well as for its positioning along the
$K$-axis.

Note that the lone stability tongue of Figure~\ref{fig:stablarge1}(d)
narrows as $\rho$ increases, eventually vanishing (at our resolution
in $K$) once $\rho$ exceeds $~0.04$. A similar, though less
pronounced, narrowing of the left-most stability tongue, with
increasing $\rho$, is present in the other stability diagrams
presented in Figure~\ref{fig:stablarge1}. We conjecture that this
narrowing is related to the observation above that these stability
tongues are associated with values of $K$ that are unstable in the
absence of feedback to spatially resonant perturbation wavenumbers
$\widetilde Q_n$.  To understand this claim, expand the real and
imaginary parts of the eigenvalue $j_{1}(Q)$ of~(\ref{eq:Jmatrix})
about the resonant $Q=\widetilde Q_n=nK$ ($n=2$ above), and focus on
the specific contribution to $j_{1}(Q)$ that is due to the spatial
feedback term. Taylor expanding the diagonal factor $\rho(e^{iQ\Delta
  x}-1)$, with $Q=\widetilde Q_n+\Delta Q$, about $\Delta Q=0$, we
find that the effect of the spatial feedback term is to move the
eigenvalue $j_1$ in the complex plane by an amount $\rho(i\Delta
Q\Delta x-\Delta Q^2\Delta x^2/2+\cdots$). For large enough $\rho$ we
thus expect the width of the band of unstable wavenumbers about
$\widetilde Q_n$ to scale as $1/\sqrt{\rho}$ since the spatial
feedback will decrease the growth rate $f(Q)$ near $\widetilde Q_n$ by
$\rho\Delta Q^2\Delta x^2/2$; {\it i.e.} the larger $\rho$ is, the
smaller the unstable band about $\widetilde Q_n$ will be.  This in
turn leads to an $\cal{O}(\sqrt{\rho})$ contribution to the change in
$j_{1i}$ over the same band of unstable wavenumbers since the spatial
feedback shifts $j_{1i}$ by $\rho\Delta Q\Delta x$.  From
Lemma~\ref{tm:degeneracies} it follows that if the change in $j_{1i}
\Delta t$ over the unstable band exceeds $2\pi$ then the traveling
wave is necessarily unstable. Thus, we expect the stable regions
associated with spatially resonant $Q$ must vanish for large enough
$\rho$.

\begin{figure}
\centerline{\resizebox{4 in}{!}{\includegraphics{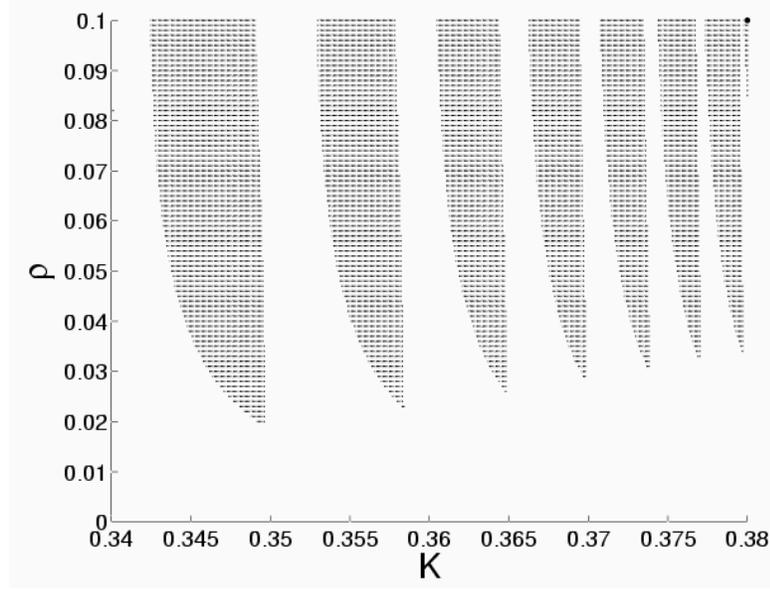}}}
\centerline{$\qquad$ (a) }
\centerline{\resizebox{4 in}{!}{\includegraphics{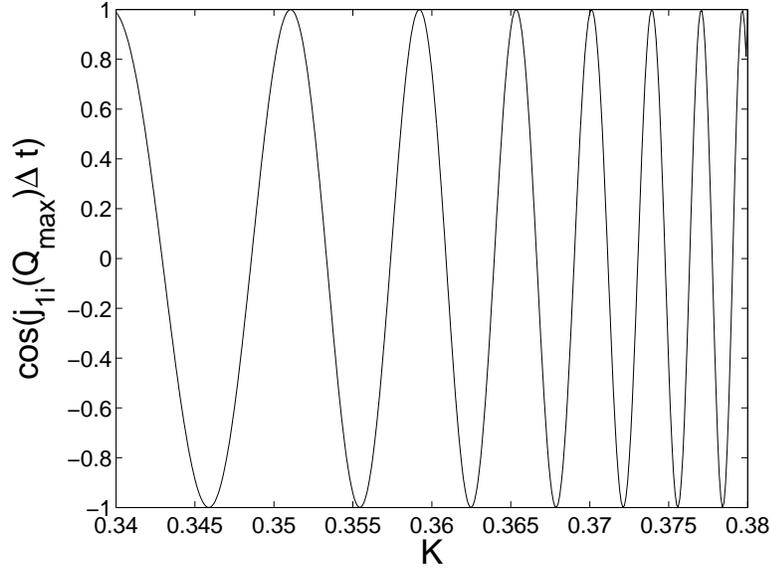}}}
\centerline{$\qquad$ (b) }
\caption{(a) A blow-up of a 
  portion of the stability diagram for $(b_1,b_3)=(2.5,2.0)$ ({\it cf.} 
Fig.~\ref{fig:stablarge1}(g)).  (b)
  Corresponding plot of $\cos(j_{1i}(Q_{max}) \Delta t)$ {\it vs.} $K$,
  where $Q_{max}(K)$ is the perturbation wavenumber that gives the
  largest growth rate $f_{max}$ for each $K$.}
\label{fig:fingers}
\end{figure}

We now focus on the accumulation of narrow stability tongues that are
evident, for example, in Figures~\ref{fig:stablarge1}(e)-(g).  We find
that the stability tongues become narrower and narrower, shifting to
larger and larger values of $\rho$, as $|K|\to \frac{1}{\sqrt{1 +
b_{1}b_{3}}}$, which is where the frequency $\omega\rightarrow 0$ and
consequently $\Delta t$ diverges.  Figure~\ref{fig:fingers}(a) shows a
blowup of part of Fig.~\ref{fig:stablarge1}(g).  The rapid alternation
between stable and unstable regions with small changes in $K$, occurs
as $K$ approaches the singular point where $\omega\to 0$ and $\Delta
t\to \infty$.  We trace this rapid variation in stability to the rapid
winding of the phase $j_{1i}(Q)\Delta t$, through successive multiples
of $2\pi$, when $\Delta t$ is large. This claim is explored in
Figure~\ref{fig:fingers}(b). Specifically, for this figure the value
of $\rho$ is held fixed at $\rho=0.1$, and the value of $Q_{max}$ as a
function of $K$ is determined; here $Q_{max}$ corresponds to the value
of $Q$ at which the growth rate $f(Q)$ reaches its absolute maximum
$f_{max}$. Figure~\ref{fig:fingers}(b) is a plot of
$\cos(j_{1i}(Q_{max})\Delta t)$ {\it vs.} $K$, which shows that the
spacing of the stability tongues in Figure~\ref{fig:fingers}(a) is the
same as the spacing of the maxima of $\cos(j_{1i}(Q_{max})\Delta t)$
in Figure~\ref{fig:fingers}(b). On closer examination, we find that
the positions of the stability tongues in Figure~\ref{fig:fingers}(a)
for $\rho = 0.1$ are approximately centered around the points for
which $j_{1i}(Q_{max}) \Delta t =(2n + 1) \pi$, for successive
integers $n$.  That the locations of the stability tongues appear to
coincide with the places where $j_{1i}(Q_{max})\Delta t=(2n+1)\pi$ is
perhaps not surprising.  In particular, as already noted, it is at
this degenerate value that the stability boundary in the
$(j_{1i},\gamma)$ parameter plane extends the furthest from the
asymptote $\gamma=j_{1r}$, and for this reason it accommodates the
largest value of $f_{max}$ stably ({\it cf.}
Fig.~\ref{fig:degenerate}(a)). (See also Just~{\it et
al.}~\cite{JBORB97}.) Moreover, we know that points for which
$j_{1i}(Q_{max}) \Delta t=2n\pi$ cannot be stable by
Lemma~\ref{tm:degeneracies}, so it is expected that the stability
tongues, if present, would avoid the even multiples of $\pi$.  It is
interesting to note that the spatial length scale for the alternation
between stable and unstable regions observed in
Figure~\ref{fig:fingers}(a) is a direct result of the introduction of
time delay into the problem.  When no time delay is included, the
stability of the traveling wave depends only upon $f(Q)$, the largest
growth rate associated with an eigenvalue of the stability matrix $J$,
but once the time delay is added both the real and imaginary parts of
$j_{1}$ can influence stability in a complicated manner captured by
our main stability result, Theorem~\ref{tm:characteristic}.

\begin{figure}
\centerline{$\qquad$ (a): $(m,n)=(1,2)$  \hspace{1.3in} (b): $(m,n)=(1,3)$ }
\smallskip
\centerline{\resizebox{3in}{!}{\includegraphics{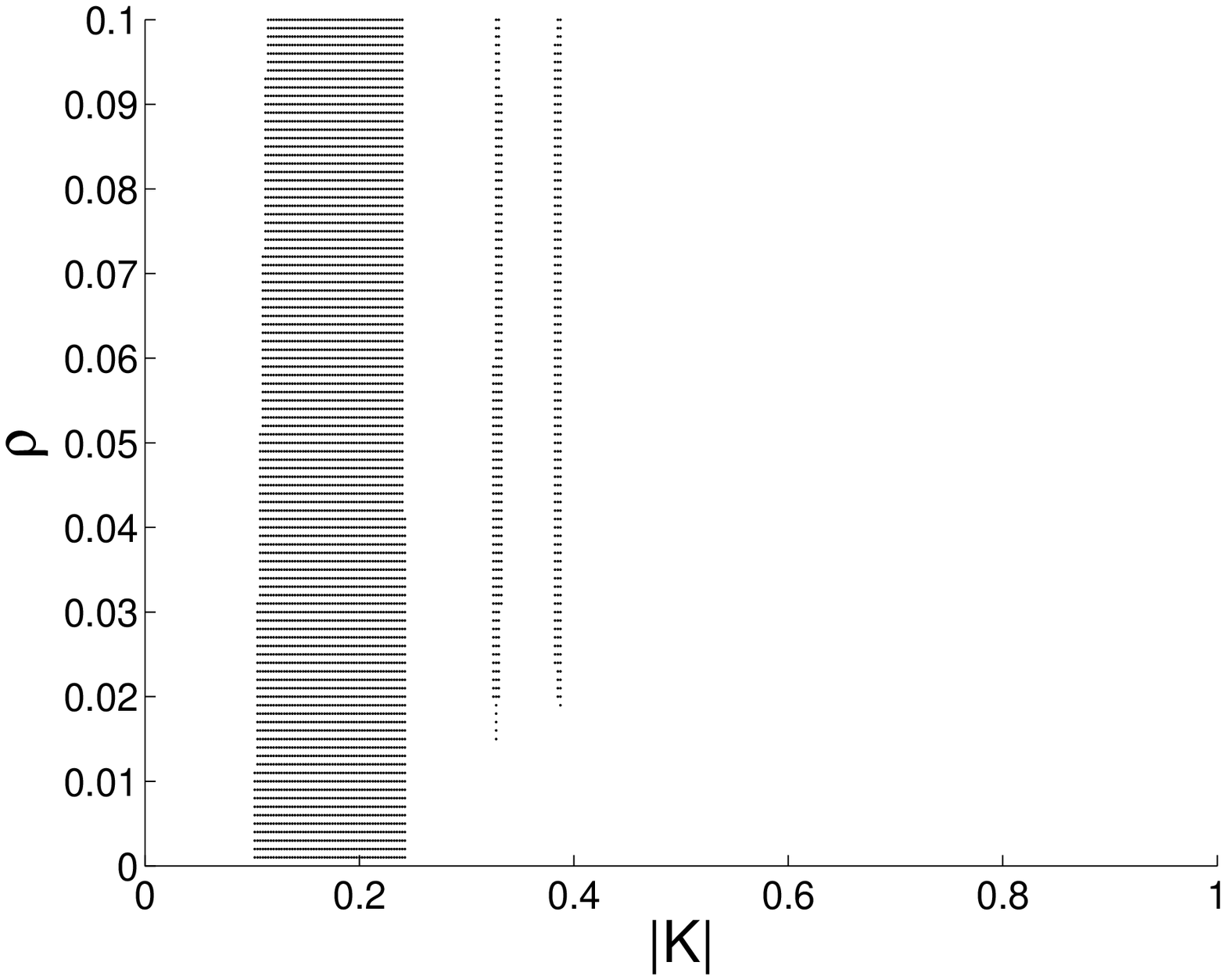}}
\resizebox{3in}{!}{\includegraphics{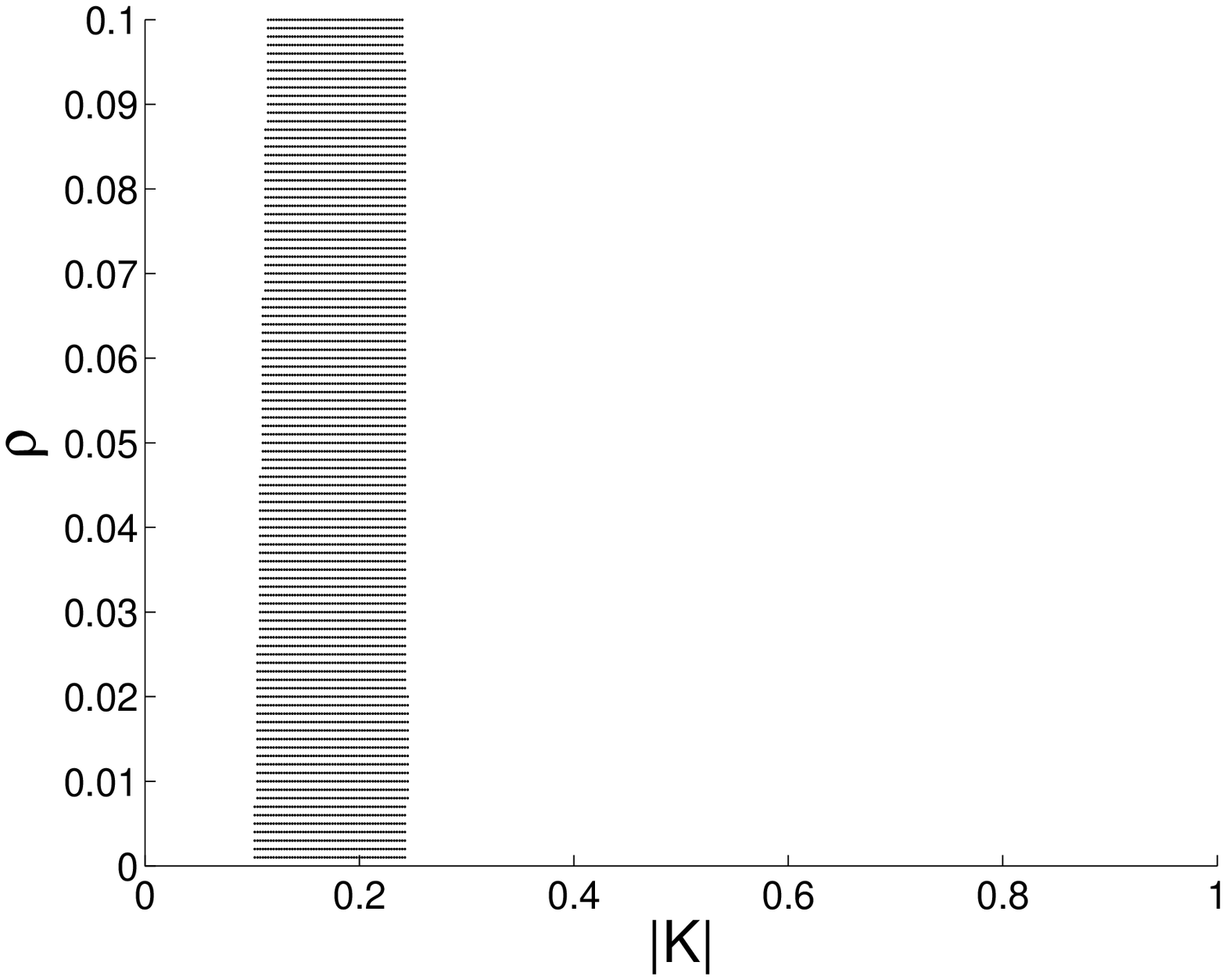}}}
\medskip
\centerline{$\qquad$ (c): $(m,n)=(2,1)$  \hspace{1.3in} (d): $(m,n)=(3,1)$ }
\smallskip
\centerline{\resizebox{3in}{!}{\includegraphics{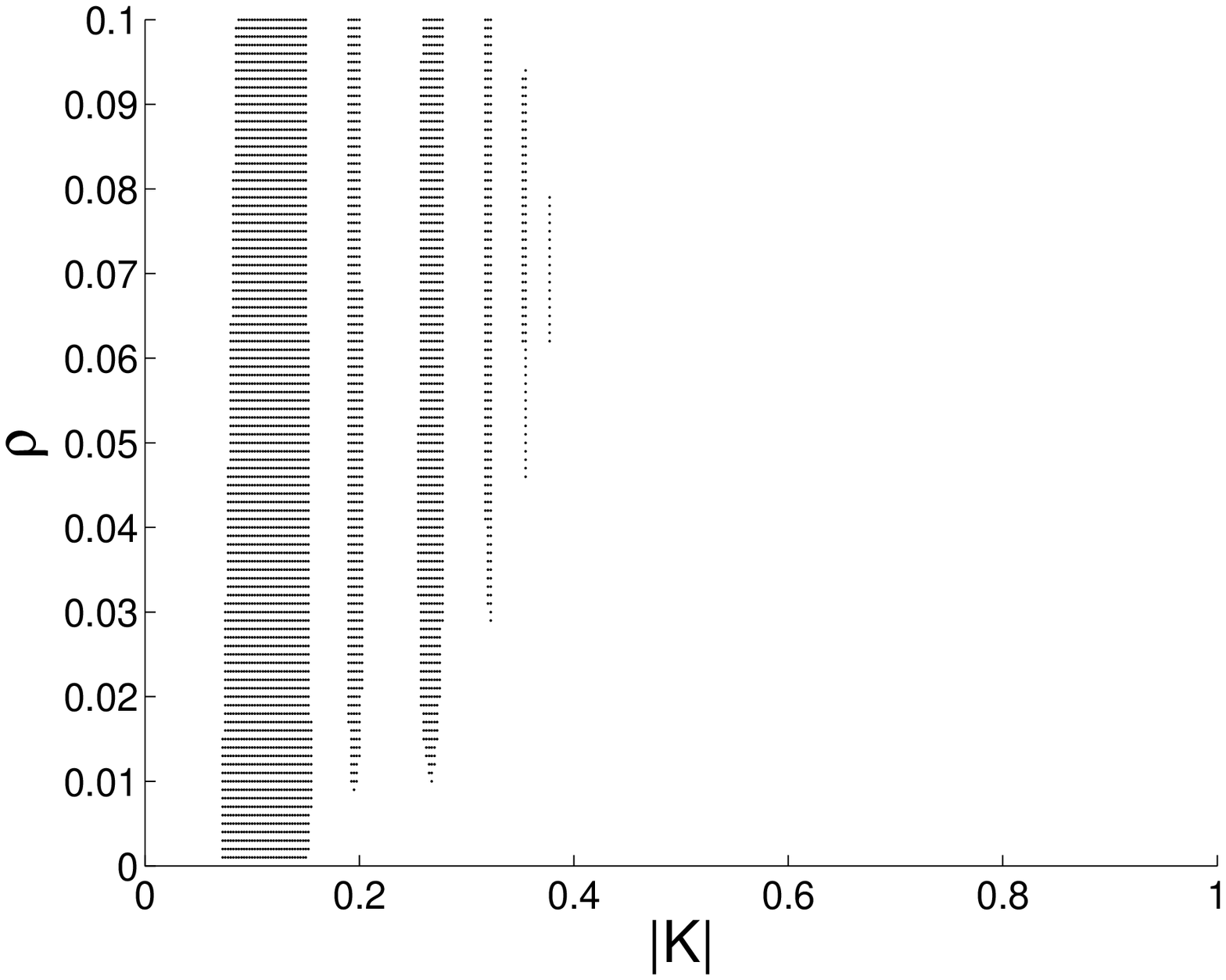}}
\resizebox{3in}{!}{\includegraphics{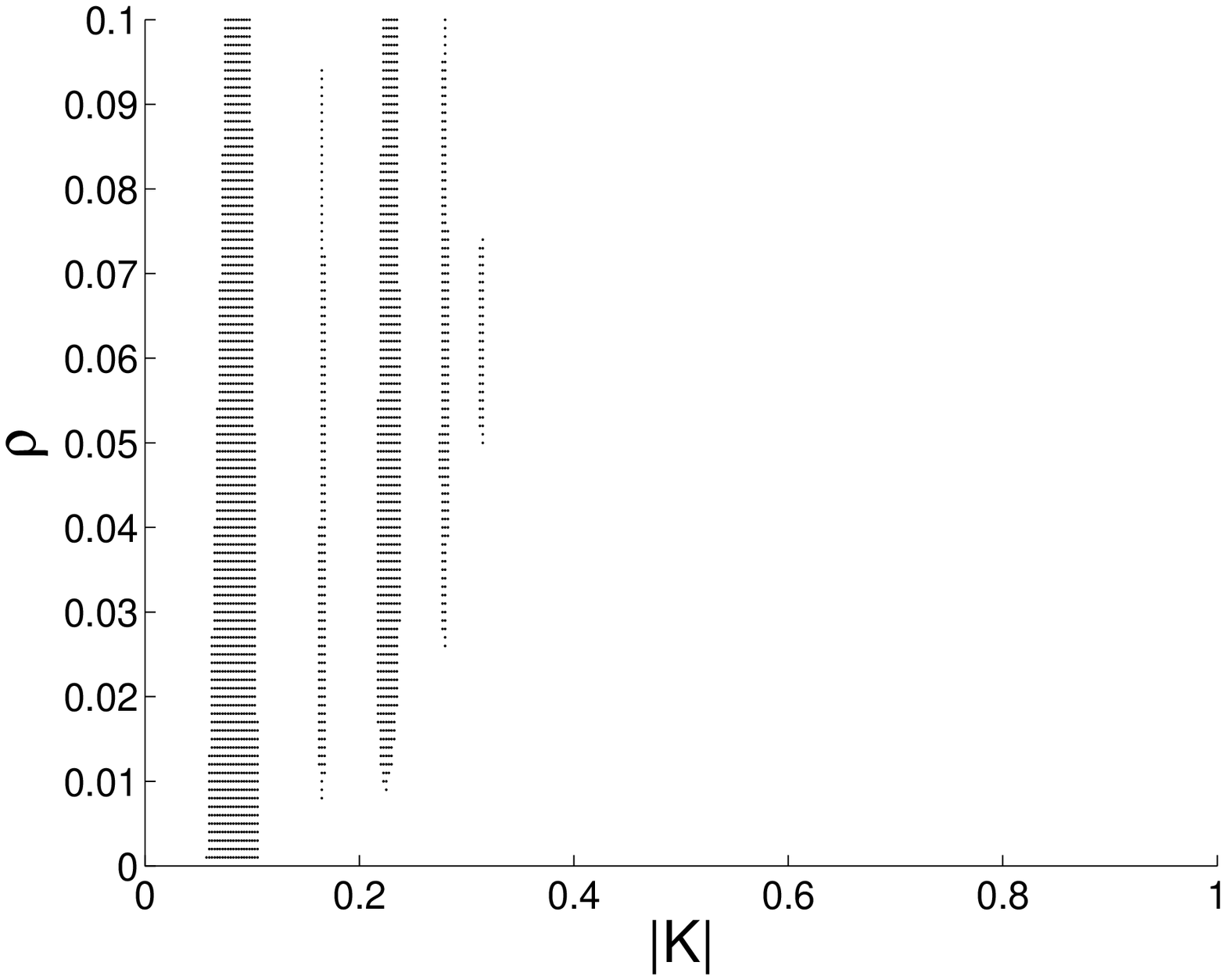}}}
\caption{Examples of stability diagrams for $(b_1,b_3)=(2.0,1.5)$ and
$\omega \Delta t = 2 m \pi$, $K \Delta x=2n\pi$,  where $(m,n)$ are
indicated above each figure. The stability diagram associated with the
standard case $(m,n)=(1,1)$ is presented in
Fig.~\ref{fig:stablarge1}(e). }
\label{fig:mnplots}
\end{figure}

While the presentation of our analysis focused on the case where
$\Delta x$ and $\Delta t$ are chosen so that $K\Delta x=|\omega|\Delta
t=2\pi$, it is readily extended to the situation where $\omega\Delta
t=2m\pi$ and $K\Delta x=2n\pi$ for nonzero integers $m$ and $n$.
Figure~\ref{fig:mnplots} presents numerical linear stability results
obtained for various integers $m,n$. It is perhaps not surprising that
the effect of increasing the delay time or the spatial shift is to
reduce the extent of the stability regions in the $(|K|,\rho)$--plane.
Afterall, the larger the value of $\Delta t$, the more likely it is
that $j_{1i}\Delta t$ will reach a multiple of $2\pi$ over an unstable
band of $Q$--values. Similarly, the larger the value of $\Delta x$,
the more likely that $Q\Delta x$ is a multiple of $2\pi$ over an
unstable interval of perturbation wavenumbers. However, increasing the
spatial shift by the factor $n$ does not appear to significantly
affect the first stability tongue (compare Fig.~\ref{fig:mnplots} (a)
and (b) with Fig.~\ref{fig:stablarge1}(e)).  This observation is
consistent with our association of this first stability tongue, which
extends all the way to $\rho=0$ with values of $Q$ for which $Q\Delta
x$ is a multiple of $2\pi$; the stabilization achieved over that band
of $K$ is due to the temporal feedback.  Increasing the time delay by
the factor $m$ has a much more significant impact on this first
stability tongue (see Fig.~\ref{fig:mnplots}(c) and (d)).  It both
narrows the stability tongues and appears to shift them to smaller
$|K|$. This effect is presumably related to the overall increase, with
$\Delta t$, in the winding of the phase $j_{1i}(Q_{max})\Delta t$ over
an interval of $K$ values, thereby accounting for the shift in the
positions of the stability tongues ({\it cf.} Fig.~\ref{fig:fingers}).

\section{Conclusions}
\label{sec:conclude}

In this paper we have examined the effectiveness of the noninvasive
feedback control scheme, proposed first in the setting of a nonlinear
optics problem~\cite{LYH96}, for stabilizing traveling wave solutions
of the one-dimensional complex Ginzburg Landau equation. Our approach
is based on a linear stability analysis, which involves examining the
solutions of a linear (complex) delay equation whose coefficients
depend on the perturbation wavenumber $Q$. The traveling wave
solution is stabilized by the feedback if all solutions of the
$Q$-parameterized family of delay equations decay.  Our analysis leads
to a single stability criterion, Theorem~\ref{tm:characteristic}, that
allows us to determine stability of a traveling wave against
perturbations of all wave numbers by determining whether a nonlinear
algebraic equation for $Q$ possesses a solution or not. Perhaps more
significant is that our main stability theorem specifies a value of
the feedback parameter $\gamma$, namely $\gamma=-\frac{1}{\Delta t}$,
which is guaranteed to work, if stabilization is at all possible.
Thus, loosely speaking, one of the ``control knobs'' can be set in
advance, thereby minimizing the amount of control parameter space that
need be scanned in trying to stabilize traveling wave solutions.

In addition to our necessary and sufficient stability criterion of
Theorem~\ref{tm:characteristic} we determined simple sufficient
conditions under which the control scheme will be ineffective and
interpreted these in terms of resonance conditions. For instance, we
find that if there is an unstable wavenumber $Q$ that is an integer
multiple of the wavenumber $K$ of the targeted solution, then the
spatially-translated feedback cannot eliminate this instability.
Likewise if there is an unstable wavenumber and the frequency
associated with that perturbation, $j_{1i}(Q)$, is an integer multiple
of the frequency $\omega$ of the underlying traveling wave state, then
the temporal feedback is ineffective in suppressing the
instability~\cite{JBORB97}. A simple consequence of these observations
is that it is not possible to use the temporal feedback to stabilize
the spatially uniform oscillatory pattern at $K=0$; this follows quite
generally from the fact that the Benjamin-Feir instability is a
longwave instability of a translation invariant problem. Specifically,
translation symmetry forces a neutral mode to exist at $Q=0$ and the
longwave nature of the instability ensures that there are
instabilities in a neighborhood of $Q=0$. For $K=0$, and $Q$
sufficiently small, these instabilities are associated with purely
real eigenvalues of the linear stability matrix $J$.  Such steady
modes of instability cannot be suppressed by spatially-translated
feedback if $K=0$, and they cannot be eliminated by the temporal
feedback either since $j_{1i}(Q)=0$ if the eigenvalues are purely
real. This observation is interesting in light of similar results
obtained by Harrington and Socolar~\cite{HS01}; building on results
in~\cite{N97,NU98} they show that the extended time-delay
autosynchronization approach to controlling traveling waves of the
CGLE in one-dimension~\cite{BS962} necessarily fails in two-dimensions
due to symmetries that force the Floquet multipliers to be purely
real.

For various values of $(b_1,b_3)$ in the Benjamin-Feir unstable
regime, we have used our stability criterion to determine numerically
where in the $(|K|,\rho)$--parameter plane the feedback control scheme
is effective in stabilizing traveling wave solutions with wavenumber
$K$. We find that the stability regions take the form of stability
tongues, and we offer some insights into the position and spacing with
$|K|$ of these stable regions, as well as their extent in the feedback
parameter $\rho$. While we find some limited regions where temporal
feedback alone works ($\rho=0)$ and another case where spatial
feedback alone works ($\gamma=0$), most stable regions require a
combination of spatial and temporal feedback to stabilize the
traveling wave. However, a notable feature of the stability diagrams
in Figure~\ref{fig:stablarge1}, is that the minimal $\rho$ required to
stabilize the waves is often quite small.

We expect Theorem~\ref{tm:characteristic} will have applications
beyond the example considered here of feedback control within the
setting of the complex Ginzburg Landau equation.  In particular, we
expect our analysis to apply to other linear stability problems that
lead to a parameterized family of delay equations of the
form~(\ref{eq:deleq1}), provided that the growth rates $j_{kr}(Q)$, in
the absence of temporal feedback, have an absolute maximum as a
function of $Q$.  Our analysis of the CGLE with feedback was further
simplified by the fact that we could eliminate one of the delay
equations from consideration altogether. We did not, however, consider
the case where the feedback parameters $\gamma$ and $\rho$ are allowed
to be complex.

This paper suggests a number of directions for further investigation.
For example, it would be interesting to carry out numerical
investigations to determine the evolution of the system in the
parameter regimes where the traveling waves are not stabilized, {\it
i.e.} choose $\Delta x$ and $\Delta t$ appropriate for a traveling
wave of wavenumber $K$ that is not stabilized by the feedback. Would
the system then evolve to a different traveling wave state, one for
which the feedback does not vanish?  More generally, numerical
investigations, guided by our analysis, could help assess how well the
feedback works when stabilization is possible, but when it is not
possible to prepare the system in a neighborhood of the targeted
traveling wave pattern.  For instance, will the feedback work when it
is ``turned on'' after a spatio-temporal chaotic state has developed?
Another avenue for numerical investigation stems from a potential
drawback associated with the feedback scheme investigated here -- it
assumes that the traveling wave dispersion relation $\omega(K)$ is
known. Hence it would be worthwhile to investigate numerically the
consequences of a ``mismatch'' between the spatial shift $\Delta x$
and the temporal delay $\Delta t$. This might be particularly
interesting in the case of the two--dimensional CGLE, in which case
the traveling wave pattern can ``rotate'' in order to compensate for
any mismatch.

Another direction for future research is suggested by our choice of
the CGLE for the detailed analysis presented here.  Specifically we
refer to the fact that the CGLE arises as the universal amplitude
equation describing the {\it long} spatial and {\it slow} temporal
evolution of homogeneous oscillations in spatially extended isotropic
systems. Are these the appropriate scales on which to be applying the
feedback? If so, how should the feedback parameters for the ``original
problem'' be chosen to achieve the optimal value of $\gamma=-1/\Delta
t$ that we find for the ``reduced model'' given by the CGLE? As a
first step in addressing these questions it would be worthwhile to
carry out an analysis, similar to the one we have done here, but on
the original equations describing the physical system undergoing a
Hopf bifurcation, {\it e.g.} on a model of chemical
reaction--diffusion system in the oscillatory regime.

An attractive feature of the feedback control scheme investigated here
is its noninvasive nature, {\it i.e.} the feedback vanishes when the
targeted state is reached. This design feature exploits the underlying
spatial and spatio-temporal symmetry properties of the desired
pattern in a very natural way, suggesting that this feedback approach
may be readily generalized to more complicated patterns. For instance,
the original problem investigated by Lu, Yu, and Harrison~\cite{LYH96}
involved oscillatory stripe patterns in a {\it two-dimensional}
isotropic system.  They found that they could target a particular
orientation of the stabilized stripe pattern by an appropriate
rotation of the spatial translation applied in the feedback control.
Thus it would be of interest both to extend our linear stability
analysis to the control of traveling plane wave solutions of the {\it
  two-dimensional} CGLE, and to consider the feedback stabilization of
more complicated traveling wave patterns.

\section*{Acknowledgments} 
We thank Sue Ann Campbell and Josh Socolar for helpful discussions
regarding this work. The research of MS was supported by NSF grant
DMS-9972059 and by the NSF MRSEC Program under DMR-0213745. KAM
received support through an NSF-IGERT fellowship under NSF grant
DGE-9987577.

\section*{References}
\bibliographystyle{unsrt}
\bibliography{mont0807}

\begin{thebibliography}{10}

\bibitem{Pyr92}
K.~Pyragas.
\newblock Continuous control of chaos by self-controlling feedback.
\newblock {\em Physics Letters A}, 170:421, 1992.

\bibitem{LYH96}
W.~Lu, D.~Yu, and R.~G. Harrison.
\newblock Control of patterns in spatiotemporal chaos in optics.
\newblock {\em Phys. Rev. Lett.}, 76:3316, 1996.

\bibitem{GSS88}
M~Golubitsky, I~Stewart, and D.G. Shaeffer.
\newblock {\em Singularities and Groups in Bifurcation Theory, Volume II}.
\newblock Springer-Verlag, New York, 1988.

\bibitem{SPSHCH92}
B.~I. Shraiman, A.~Pumir, W.~van Saarloos, P.~C. Hohenberg, H.~Chat\'{e}, and
  M.~Holen.
\newblock Spatiotemporal chaos in the one-dimensional complex
  {G}inzburg-{L}andau equation.
\newblock {\em Physica D}, 57:241, 1992.

\bibitem{OGY90}
E.~Ott, C.~Grebogi, and J.~A. Yorke.
\newblock Controlling chaos.
\newblock {\em Phys. Rev. Lett.}, 64:1196, 1990.

\bibitem{PT93}
K.~Pyragas and A.~Tama\v{s}evi\v{c}ius.
\newblock Experimental control of chaos by delayed self-controlling feedback.
\newblock {\em Physics Letters A}, 180:99, 1993.

\bibitem{GSCS94}
D.~J. Gauthier, D.~W. Sukow, H.~M. Concannon, and J.~E.~S. Socolar.
\newblock Stabilizing unstable periodic orbits in a fast diode resonator using
  continuous time-delay autosynchronization.
\newblock {\em Phys. Rev. E}, 50:2343, 1994.

\bibitem{BDG94}
S.~Bielawski, D.~Derozier, and P.~Glorieux.
\newblock Controlling unstable periodic orbits by a delayed continuous
  feedback.
\newblock {\em Phys. Rev. E}, 49:R971, 1994.

\bibitem{PBA96}
Th. Pierre, G.~Bonhomme, and A.~Atipo.
\newblock Controlling the chaotic regime of nonlinear ionization waves using
  the time-delay autosynchronization method.
\newblock {\em Phys. Rev. Lett.}, 76:2290, 1996.

\bibitem{MKPAPB97}
Th. Mausbach, Th. Klinger, A.~Piel, A.~Atipo, Th. Pierre, and G.~Bonhomme.
\newblock Continuous control of ionization wave chaos by spatially derived
  feedback signals.
\newblock {\em Physics Letters A}, 228:373, 1997.

\bibitem{FSK02}
T.~Fukuyama, H.~Shirahama, and Y.~Kawai.
\newblock Dynamical control of the chaotic state of the current-driven ion
  acoustic instability in a laboratory plasma using delayed feedback.
\newblock {\em Physics of Plasmas}, 9:4525, 2002.

\bibitem{SBFHLM93}
F.~W. Schneider, R.~Blittersdorf, A.~F\"{o}rster, T.Hauck, D.~Lebender, and
  J.~M\"{u}ller.
\newblock Continuous control of chemical chaos by time delayed feedback.
\newblock {\em J. Phys. Chem.}, 97:12244, 1993.

\bibitem{LFS95}
A.~Lekebusch, A.~F\"{o}rster, and F.~W. Schneider.
\newblock Chaos control in an enzymatic reaction.
\newblock {\em J. Phys. Chem.}, 99:681, 1995.

\bibitem{PMRNKG}
P.~Parmananda, R.~Madrigal, M.~Rivera, L.~Nyikos, I.~Z. Kiss, and
  V.~G\'asp\'ar.
\newblock Stabilization of unstable steady states and periodic orbits in an
  electrochemical system using delayed-feedback control.
\newblock {\em Phys. Rev. E}, 59:5266, 1999.

\bibitem{LCM98}
A.~Labate, M.~Ciofini, and R.~Meucci.
\newblock Controlling quasiperiodicity in a {CO}$_2$ laser with delayed
  feedback.
\newblock {\em Phys. Rev. E}, 57:5230, 1998.

\bibitem{SSG94}
J.E.S. Socolar, D.W. Sukow, and D.J. Gauthier.
\newblock Stabilizing unstable periodic orbits in fast dynamical systems.
\newblock {\em Phys. Rev. E}, 50:3245, 1994.

\bibitem{BS962}
M.~E. Bleich and J.~E.~S. Socolar.
\newblock Controlling spatiotemporal dynamics with time-delay feedback.
\newblock {\em Phys. Rev. E}, 54:R17, 1996.

\bibitem{HS01}
I.~Harrington and J.E.S. Socolar.
\newblock Limitations on stabilizing plane waves via time-delay feedback.
\newblock {\em Phys. Rev. E}, 64:056206, 2001.

\bibitem{BHMS97}
M.~E. Bleich`, D.~Hochheiser, J.~V. Moloney, and J.~E.~S. Socolar.
\newblock Controlling extended systems with spatially filtered, time-delayed
  feedback.
\newblock {\em Phys. Rev. E}, 55:2119, 1997.

\bibitem{BASJ02}
N.~Baba, A.~Amann, E.~Sch\"{o}ll, and W.~Just.
\newblock Giant improvement of time-delayed feedback control by spatio-temporal
  filtering.
\newblock {\em Phys. Rev. Lett.}, 89:074101, 2002.

\bibitem{DV96}
E.~V. Degtiarev and M.~A. Vorontsov.
\newblock Dodecagonal patterns in a {K}err-slice/feedback-mirror type optical
  system.
\newblock {\em Journal of Modern Optics}, 43:93, 1996.

\bibitem{MSOF98}
R.~Martin, A.~J. Scroggie, G.-L. Oppo, and W.~J. Firth.
\newblock Stabilization, selection, and tracking of unstable patterns by
  {F}ourier space techniques.
\newblock {\em Phys. Rev. Lett.}, 77:4007, 1996.

\bibitem{MS98}
A.~V. Mamaev and M.~Saffman.
\newblock Selection of unstable patterns and control of optical turbulence by
  {F}ourier plane filtering.
\newblock {\em Phys. Rev. Lett.}, 80:3499, 1998.

\bibitem{VS98}
M.~A. Vorontsov and B.~A. Samson.
\newblock Nonlinear dynamics in an optical system with controlled
  two-dimensional feedback: Black-eye patterns and related phenomenon.
\newblock {\em Phys. Rev. A}, 57:3040, 1998.

\bibitem{BKNT00}
E.~Benkler, M.~Kreuzer, R.~Neubecker, and T.~Tshudi.
\newblock Experimental control of unstable patterns and elimination of
  spatiotemporal disorder in nonlinear optics.
\newblock {\em Phys. Rev. Lett.}, 84:879, 2000.

\bibitem{FBS99}
G.~Franceschini, S.~Bose, and E.~Sch\"oll.
\newblock Control of chaotic spatiotemporal spiking by time-delay
  autosynchronization.
\newblock {\em Phys. Rev. E}, 60:5426, 1999.

\bibitem{KBPvMRE01}
M.~Kim, M.~Bertram, M.~Pollmann, A.~von Oertzen, A.S. Mikhailov, H.H.
  Rotermund, and G.~Ertl.
\newblock Controlling chemical turbulence by global delayed feedback: Pattern
  formation in catalytic {CO} oxidation on {P}t(110).
\newblock {\em Science}, 292:1357, 2001.

\bibitem{BM01}
M.~Bertram and A.S. Mikhailov.
\newblock Pattern formation in a surface chemical reaction with global delayed
  feedback.
\newblock {\em Phys. Rev. E}, 63:066102, 2001.

\bibitem{BBMRE03}
C.~Beta, M.~Bertram, A.S. Mikhailov, H.H. Rotermund, and G.~Ertl.
\newblock Controlling turbulence in a surface chemical reaction by time-delay
  autosynchronization.
\newblock {\em Phys. Rev. E}, 67:046224, 2003.

\bibitem{BM96}
D.~Battogtokh and A.~Mikhailov.
\newblock Controlling turbulence in the complex {G}inzburg-{L}andau equation.
\newblock {\em Physica D}, 90:84, 1996.

\bibitem{BPM97}
D.~Battogtokh, A.~Preusser, and A.~Mikhailov.
\newblock Controlling turbulence in the complex {G}inzburg-{L}andau equation
  {II}. two-dimensional systems.
\newblock {\em Physica D}, 106:327, 1997.

\bibitem{JBORB97}
W.~Just, T.~Bernard, M.~Ostheimer, E.~Reibold, and H.~Benner.
\newblock Mechanism of time-delayed feedback control.
\newblock {\em Phys. Rev. Lett.}, 78:203, 1997.

\bibitem{N97}
H.~Nakajima.
\newblock On analytical properties of delayed feedback control of chaos.
\newblock {\em Phys. Lett. A}, 232:207, 1997.

\bibitem{NU98}
H.~Nakajima and Y.~Ueda.
\newblock Limitation of generalized delayed feedback control.
\newblock {\em Physica D}, 111:143, 1998.

\bibitem{P01}
K.~Pyragas.
\newblock Control of chaos via an unstable delayed feedback controller.
\newblock {\em Phys. Rev. Lett.}, 86:2265, 2001.

\bibitem{JRKFHB00}
W.~Just, E.~Reibold, K.~Kacperski, P.~Fronczak, J.~A. Holyst, and H.~Benner.
\newblock Influence of stable {F}loquet exponents on time-delayed feedback
  control.
\newblock {\em Phys. Rev. E}, 61:5045, 2000.

\bibitem{JBR03}
W.~Just, H.~Benner, and E.~Reibold.
\newblock Theoretical and experimental aspects of chaos control by time-delayed
  feedback.
\newblock {\em Chaos}, 13:259, 2003.

\bibitem{P02}
K.~Pyragas.
\newblock Analytical properties and optimization of time-delayed feedback
  control.
\newblock {\em Phys. Rev. E}, 66:026207, 2002.

\bibitem{BS96}
M.~E. Bleich and J.~E.~S. Socolar.
\newblock Stability of periodic orbits controlled by time-delay feedback.
\newblock {\em Physics Letters A}, 210:87, 1996.

\bibitem{RSJ2000}
D.~V.~Ramana Reddy, A.~Sen, and G.~L. Johnston.
\newblock Dynamics of a limit cycle oscillator under time delayed linear and
  nonlinear feedbacks.
\newblock {\em Physica D}, 144:355, 2000.

\bibitem{BF67}
T.B. Benjamin and J.E. Feir.
\newblock The disintegration of wave trains on deep water {P}art 1. {T}heory.
\newblock {\em J. Fluid Mech.}, 27:417, 1967.

\bibitem{SD78}
J.T. Stuart and R.C. DiPrima.
\newblock The {E}ckhaus and {B}enjamin-{F}eir resonance mechanisms.
\newblock {\em Proc. R. Soc. Lond. A}, 362:27, 1978.

\bibitem{JPBCRK92}
B.~Janiaud, A.~Pumir, D.~Bensimon, V.~Croquette, H.~Richter, and L.~Kramer.
\newblock The {E}ckhaus instability for traveling waves.
\newblock {\em Physica D}, 55:269, 1992.

\bibitem{Chate94}
H.~Chat\'{e}.
\newblock Spatiotemporal intermittency regimes of the one-dimensional complex
  {G}inzburg-{L}andau equation.
\newblock {\em Nonlinearity}, 7:185, 1994.

\bibitem{Diek}
O.~Diekmann, S.~A. van Gils, S.~M. Verduyn-Lunel, and H.~O. Walther.
\newblock {\em Delay Equations, Functional, Complex, and Nonlinear Analysis}.
\newblock Springer, New York, 1995.

\bibitem{Driver}
R.~D. Driver.
\newblock {\em Ordinary and Delay Differential Equations}.
\newblock Springer-Verlag, New York, 1997.

\end{thebibliography}
\end{document}